\documentclass[journal]{IEEEtran}
\usepackage{latexsym}
\usepackage{amsfonts,amssymb,amsmath}
\usepackage{hyperref}
\def\BibTeX{{\rm B\kern-.05em{\sc i\kern-.025em b}\kern-.08em T\kern-.1667em\lower.7ex\hbox{E}\kern-.125emX}}
\usepackage{subfigure}
\usepackage{algorithm}
\usepackage{algpseudocode}
\usepackage{amsmath}
\usepackage{amsfonts}
\usepackage{graphicx}
\usepackage{epsfig}
\usepackage{cite}
\usepackage{txfonts}
\usepackage{relsize}
\usepackage{color}
\usepackage{color}
\usepackage{CJK}
\usepackage{times}
\usepackage{multicol}
\usepackage{stfloats}
\usepackage{float}
\usepackage{lipsum}
\usepackage{bm}

\begin{document}
%Performance Analysis of RIS-Aided Cell-Free Massive MIMO Systems under Channel Aging
\title{Impact and Mitigation of Channel Aging and Electromagnetic Interference on RIS-Assisted Cell-Free Massive MIMO Systems
\thanks{This work was supported by the Research Grants Council under the Area of Excellence scheme grant AoE/E-601/22-R. This work was partly presented in IEEE International Mediterranean Conference on Communications and Networking, Madrid, Spain, July 2024. \textit{(Corresponding authors: Jun Qian; Chi Zhang.)} } }%Cell-Free Massive MIMO with Space-Constrained Access Points: The Analysis of Mutual Coupling and Spatial Correlation}
% The Effect of Mutual Coupling and Spatial Correlation on Cell-Free Massive MIMO with Space-Constrained Access Points

\author{Jun~Qian,~\IEEEmembership{Member,~IEEE,}
    Chi~Zhang,~\IEEEmembership{Student~Member,~IEEE,}
        Ross~Murch,~\IEEEmembership{Fellow,~IEEE,}
and~Khaled~B.~Letaief,~\IEEEmembership{Fellow,~IEEE}
\thanks{Jun Qian and Chi Zhang are with the Department of Electronic and Computer Engineering, The Hong Kong University of Science and Technology, Hong Kong (e-mail: eejunqian@ust.hk, czhangcc@connect.ust.hk).}
\thanks{Ross Murch and Khaled B. Letaief are with the Department of Electronic and Computer Engineering, The Hong Kong University of Science and Technology,
Hong Kong (e-mail: eermurch@ust.hk, eekhaled@ust.hk).}
%\thanks{This work was partly presented in IEEE International Mediterranean Conference on Communications and Networking, Madrid, Spain, July 2024.}
}

\maketitle
%\IEEEtitleabstractindextext
{\begin{abstract}

Cell-free massive multiple-input multiple-output (MIMO) and reconfigurable intelligent surfaces (RISs) are two potential sixth-generation (6G) technologies. However, channel aging due to user mobility and electromagnetic interference (EMI) impinging on RISs can negatively affect performance. Existing research on RIS-assisted cell-free massive MIMO systems often overlooks these issues. This work focuses on the impact and mitigation of channel aging and EMI on RIS-assisted cell-free massive MIMO systems over spatially correlated channels. To mitigate the degradation caused by these issues, we introduce a novel two-phase channel estimation scheme with large-scale fading coefficient-aided pilot assignment to enhance channel estimation accuracy compared to conventional minimum mean square error estimators. We then develop closed-form expressions for the downlink spectral efficiency (SE) performance and using these, optimize the sum downlink SE with respect to the RIS coefficient matrices. This optimization is accomplished by the projected gradient ascent (GA) algorithm. The results show that our proposed two-phase channel estimation scheme can achieve a nearly $10\%$-likely SE improvement compared to conventional channel estimation in environments affected by channel aging. A further $10\%\sim15\%$-likely SE improvement is achieved using the proposed GA algorithm compared to random RIS phases, especially when the number of RISs increases.

\end{abstract}

\begin{IEEEkeywords}
Cell-free massive MIMO, channel aging, electromagnetic interference, reconfigurable intelligent surface, spatial correlation, spectral efficiency.
\end{IEEEkeywords}}

\maketitle

\section{Introduction}
\IEEEPARstart
{I}{n} recent decades, the demand for wireless communication has surged dramatically, reflecting exponential growth. This includes the requirement for higher data rates and widespread connectivity\cite{7827017,9665300,9663100,8808168}. The increasing transmission demands have resulted in innovative technologies, such as massive multiple-input multiple-output (MIMO) systems\cite{9665300,10225319,7438738,10167480}. However, conventional cellular networks still experience severe inter-cell interference, particularly for cell-edge users limiting overall performance\cite{9779130,9665300}. 

It has been shown that cell-free massive MIMO system can provide effective inter-cell interference mitigation and accomplish quality-of-service (QoS) improvement of cell-edge users\cite{10167480,10225319,10571171,7827017}. Numerous access points (APs), geographically distributed in the cell-free massive MIMO region without cell boundaries and connected to a central processing unit (CPU), can serve users simultaneously\cite{7917284,9340389,9416909,9665300}.
APs introduce different contributions to the effective channel gain due to their geographic spread, and conjugate beamforming can satisfy the required performance with computational complexity reduction in a distributed manner\cite{7827017,9340389}. The authors in \cite{7917284,10167480} introduced a large-scale fading decoding (LSFD) process to achieve better uplink spectral efficiency (SE) than the matched filter (MF) in cell-free massive MIMO systems. Also, \cite{9064545,9840355} proposed and utilized the dynamic cooperation cluster concept to introduce scalable cell-free massive MIMO. Despite the above advantages, cell-free massive MIMO still experiences challenges in guaranteeing good service quality under harsh propagation conditions \cite{9665300,10167480}. These challenges require the development of advanced technologies to satisfy the required quality of service.

Reconfigurable intelligent surfaces (RIS) have also attracted great interest in sixth-generation (6G) networks. By
smartly shaping electromagnetic-level radio waves without active power amplifiers, RISs introduce a new degree of freedom \cite{9326394,10167480,9905943,9875036}. Integrating RISs into wireless networks introduces controllable cascaded links to assist users with undesirable channel propagation conditions \cite{9665300,10225319}. Many important characteristics of RISs have been studied. For example, the authors in \cite{9665300} introduced an optimal RIS phase shift design for channel estimation error minimisation. \cite{10225319,9765773} proposed a modified channel estimation scheme to achieve better estimation accuracy by introducing higher channel estimation overhead. 
%However, due to its inherent physical nature, an RIS cannot selectively reflect the signals impinging its surface; therefore, it also reflects undesired signals from the surrounding environment, causing electromagnetic interference (EMI) at the receiver side
{\color{black}However, besides the expected signals of the proposed system, uncontrollable wireless signals from outside sources in the physical environment can cause inevitable and non-negligible electromagnetic interference (EMI) or ‘‘pollution" impinging on RISs, with an energy proportional to the area of RIS \cite{9598875}.} Since RISs cannot selectively reflect signals,  in\cite{10134557,9598875}, EMI can also further degrade RIS-assisted system performance.
The results in \cite{9817451} with RIS-assisted ultra-reliable and low-latency wireless communication and
\cite{10133717,9598875} with RIS-assisted wireless communication further indicated that the effect of EMI is inevitable and can degrade system performance significantly. As such, the existence of EMI is uncontrollable and non-negligible in RIS-assisted systems and introduces performance limits \cite{9598875,10167480}.

Recently, researchers have focused on integrating cell-free Massive MIMO and RISs to exploit their advantages jointly\cite{9322151,9838672,10001167}. The authors in \cite{9838672} utilized hybrid active and passive RIS elements in RIS-assisted cell-free massive MIMO. \cite{10001167} studied the uplink performance of RIS-assisted cell-free massive MIMO with spatial correlation. Channel estimation might be challenging since passive RIS elements make conventional channel estimation infeasible, and increasing RIS elements increases
training overhead \cite{9875036}. Thus, \cite{10225319,9322151} introduced and modified an ON/OFF channel estimation scheme. \cite{10621117} proposed a two-phase channel estimation scheme with fractional power control-aided pilot assignment to enhance estimation accuracy. Besides \cite{10001167}, the importance of studying the RIS spatial correlation was also indicated in \cite{9665300,10264149}. %[JUN: You need to explain and define what you mean by EMI] 
Assuming EMI to have a uniform
distribution across all incident angles, the model for the effect of EMI with RIS spatial correlation has been obtained
\cite{10134557,9598875,10167480}. This shows that highly correlated RIS elements could introduce severe EMI to degrade RIS-assisted cell-free massive MIMO system performance negatively \cite{10167480}.
However, most works on RIS-assisted cell-free massive MIMO ignore the negative effect of EMI and only consider the system signals \cite{9322151,9838672,10001167}. Inspired by the study of EMI in RIS-assisted systems \cite{10133717,9598875,9817451}, the authors in \cite{10167480} first evaluated the uplink performance of RIS-assisted cell-free massive MIMO assisted by RIS with EMI. Since most existing analyses do not apply to RIS-assisted cell-free massive MIMO systems experiencing EMI, limited work has focused on this topic; investigating the performance limits to introduce system design guidelines for RIS-assisted cell-free massive MIMO systems with EMI should be focal.

Furthermore, current works of RIS-assisted cell-free massive MIMO systems are usually based on the block-fading model \cite{10167480,10225319,9665300,9779130}. However, in practice, users are not static, and the relevant user mobility will introduce continuous channel evolution, which results in the channel aging effect \cite{9416909,9875036}. The work in \cite{9416909} characterised the channel aging effect on cell-free massive MIMO system performance and compared the performance with the conventional small-cell networks. \cite{9905943,9875036,10418910} investigated the system performance of RIS-assisted massive MIMO systems with channel aging effect. Moreover, \cite{9875036,10418910} implemented the RIS coefficient matrix design to maximize the SE performance. However, studying the RIS-assisted cell-free massive MIMO system performance with channel aging remains necessary for real-world mobile transmission scenarios.

To the best of our knowledge, no previous research has examined the combined impact of channel aging and EMI on RIS-assisted cell-free massive MIMO systems, nor has the optimization of the RIS coefficient matrix been investigated in the proposed contexts. However, the interplay between channel aging and EMI is significant and warrants further exploration. Therefore, this paper is the first to develop a system model and perform a performance analysis of spatially correlated RIS-assisted cell-free massive MIMO systems considering both channel aging and EMI. We also investigate the two-phase channel estimation scheme that we have proposed in\cite{10621117} 
to a channel aging-aware environment to mitigate the effect of channel aging and EMI. Our analysis is comprehensive, where conjugate beamforming is applied to the downlink transmission, and downlink fractional power control is introduced to improve downlink SE. To improve system performance, we also propose a RIS coefficient matrix design to maximize the sum downlink SE utilizing the projected gradient ascent (GA) algorithm. The major contributions of this paper include the following:

• We develop a RIS-assisted cell-free massive MIMO system model that accounts for channel aging and EMI, incorporating spatial correlation at both the APs and RISs.

• We introduce an innovative two-phase channel estimation scheme designed to mitigate performance degradation stemming from channel aging and EMI. This approach builds on our previous work \cite{10621117}, adapting it to a channel aging-aware environment. It highlights the advantages of enhanced estimation accuracy, including pilot assignment based on large-scale fading coefficients.

• We evaluate system performance by deriving closed-form expressions for downlink SE. These closed-form expressions enable us to get important insights for system design and performance analysis of the proposed system.
Then, motivated by \cite{7031971,9875036,10418910}, we maximize the sum downlink SE with respect to the design of RIS coefficient matrices based on our derived closed-form SE expressions. We accomplish the maximization problem utilizing the projected GA algorithm.

• We present numerical results to validate our theoretical analyses and offer system design guidelines. Additionally, we examine the impact of channel aging and EMI on downlink SE. The findings demonstrate the effectiveness of the proposed two-phase channel estimation and the optimization algorithm for the RIS coefficient matrix in alleviating performance degradation. Moreover, increasing the number of APs and implementing RISs can further mitigate performance issues, particularly in scenarios with moderate EMI. However, in environments with high EMI, the full benefits of RISs may not be realized.

The remainder of the paper is structured as follows. Section II describes the spatially correlated channel model with channel aging and EMI. Section III introduces our novel two-phase channel estimation scheme, which can mitigate the effects of channel aging and EMI. Section IV derives the achievable downlink SE with fractional power control and optimizes the RIS coefficient matrices to maximize the sum downlink SE. Section V provides numerical results and insights, and Section VI summarizes the current work and proposes future work.

\section{System Model}
 
\subsection{Spatially-Correlated RIS-assisted Channel Model}

\begin{figure}[!t]
\centering
\includegraphics[width=0.72\columnwidth]{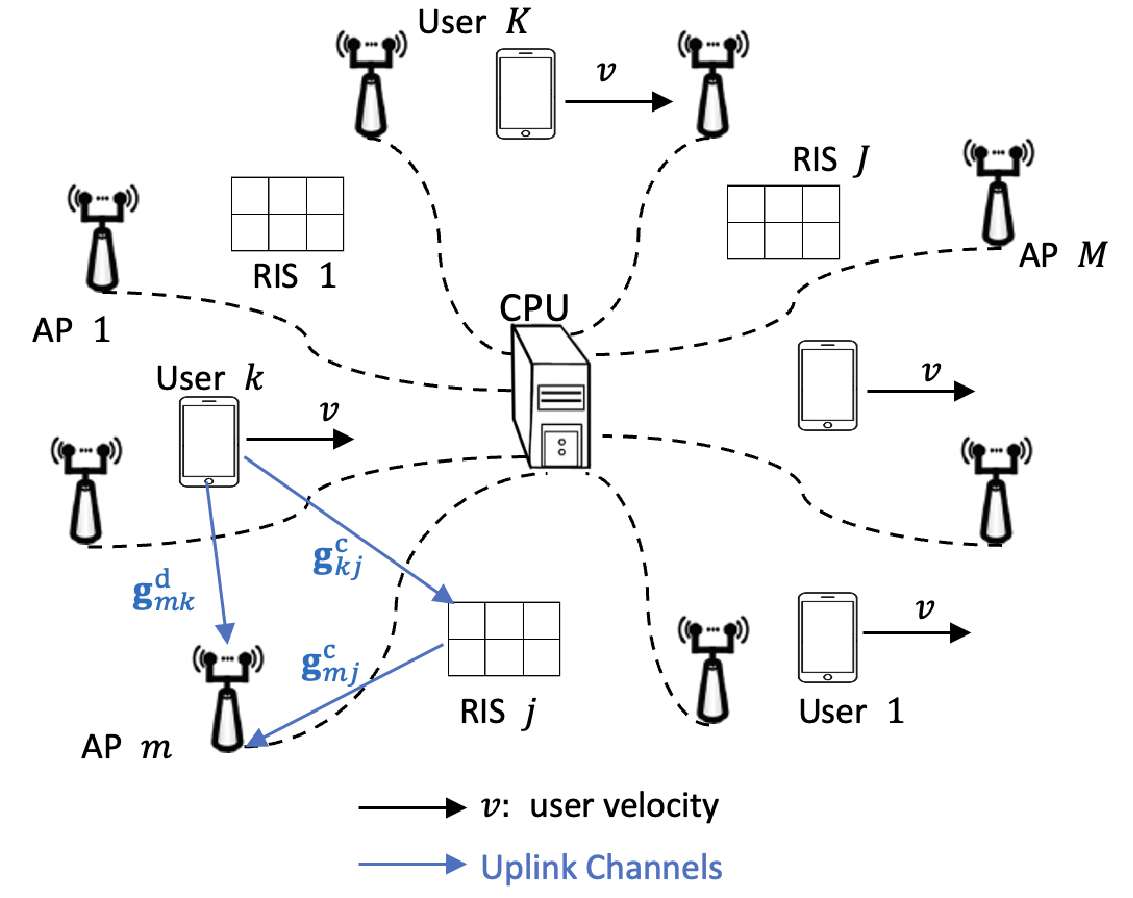}
\caption{RIS-assisted cell-free massive MIMO systems with user mobility.}
\label{Fig_1}
\vspace{-6 pt}
\end{figure}

As shown in Fig. \ref{Fig_1}, this work considers a RIS-assisted cell-free massive MIMO system operating in a time-division duplex (TDD) mode\cite{10621117}. $M$ APs serve $K$ single-antenna users simultaneously. All APs, equipped with $N$ antennas per AP, are connected to the CPU via ideal backhaul links\cite{9838672}. $J$ RISs with $L$ passive reflecting elements per RIS provide communication assistance between APs and users.
{\color{black}Note that multiple RISs will introduce multi-hop links, namely, the
transmission links from APs and reflected
by multiple RISs until reaching users\cite{10197193}, and these second-order effects are ignored in this work.}
Establishing line-of-sight paths is not possible due to user mobility and unknown obstacles and thus, the Rayleigh fading model is considered in this work\cite{9905943}. Then, the aggregate uplink channel from the $k$-th user to the $m$-th AP at time instant $n$ is modelled as \cite{10225319}
%\vspace{-6 pt}
\begin{equation}
     \displaystyle \textbf{g}_{mk}[n]=\textbf{g}_{mk}^{\text{d}}[n]+\sum\nolimits_{j=1}^{J}\textbf{g}_{mkj}^{\text{c}}[n].
     \label{aggregate_uplink_channel}
   \end{equation}
A spatially correlated Rayleigh fading model for the associated channels is considered since multi-antenna APs and multi-element RISs are deployed in our work\cite {9875036,9905943}. First, the direct channel from the $k$-th user to the $m$-th AP at time instant $n$, $\textbf{g}_{mk}^{\text{d}}[n]\in \mathbb{C}^{N\times1}$, is expressed as
\begin{equation}
     \displaystyle \textbf{g}_{mk}^{\text{d}}[n]=\sqrt{\beta_{mk}}{\textbf{R}}_{mk}^{1/2}\textbf{v}_{mk}^{\text{d}}[n],
     \label{direct_uplink_channel}
   \end{equation}
where $\beta_{mk}$ denotes the large-scale fading coefficient between the $m$-th AP and the $k$-th user, ${\textbf{R}}_{mk} \in \mathbb{C}^{N\times N}$ is the spatial correlation matrix of the $m$-th AP, $\textbf{v}_{mk}^{\text{d}}[n]\sim \mathcal{CN}(\textbf{0},\textbf{I}_{N})$ refers to the independent fast-fading channel at time instant $n$. %of the antennas 

Then, the cascaded channel from the $k$-th user to the $m$-th AP via the $j$-th RIS at time instant $n$, $\textbf{g}_{mkj}^{\text{c}}[n]$, is given by
\begin{equation}
     \displaystyle
     \textbf{g}_{mkj}^{\text{c}}[n]=\textbf{g}_{mj}^{\text{c}}[n]\boldsymbol{\Phi}_{j}[n]\textbf{g}_{kj}^{\text{c}}[n],
    \label{cascaded_uplink_channel_via_RIS}
   \end{equation}
   where 
$\textbf{g}_{mj}^{\text{c}}[n]\in \mathbb{C}^{N\times L}$, the channel from the $j$-th RIS to the $m$-th AP based on the classical Kronecker channel model introduced in \cite{7500452,9685245,qian2024performanceanalysisstarrisassistedcellfree}, is written as
\begin{equation}
     \displaystyle \textbf{g}_{mj}^{\text{c}}[n]=\sqrt{\beta_{mj}}{\textbf{R}}_{mj,r}^{1/2}\textbf{v}_{mj}^{\text{c}}[n]{\textbf{R}}_{mj,t}^{{1/2}},
     \label{channel_RIS_AP}
   \end{equation}
where $\beta_{mj}$ represents the large-scale fading coefficient between $m$-th AP and $j$-th RIS, the random variables in $\textbf{v}_{mj}^{\text{c}}[n]\in \mathbb{C}^{N\times L}$ are independent and identically distributed (i.i.d.) and satisfying $\mathcal{N}_{c}(0,1)$. ${\textbf{R}}_{mj,r} \in \mathbb{C}^{N\times N}$ and ${\textbf{R}}_{mj,t}=A_j{\textbf{R}_j} \in \mathbb{C}^{L\times L}$ \cite{9598875,9300189} are the one-sided correlations at $m$-th AP and $j$-th RIS, respectively. $A_j=d_Vd_H$ is the RIS element area; the vertical height and the horizontal width are $d_V$ and $d_H$, respectively \cite{9300189}. The $(x,y)$-th element in $\textbf{R}_j$ is formulated as \cite{9598875}\cite{9300189}
 \begin{equation}
     \begin{array}{c@{\quad}c}
\displaystyle [\textbf{R}_j]_{x,y}=\text{sinc}\Bigg{(}\frac{2||\textbf{u}_x-\textbf{u}_y||}{\lambda_c}\Bigg{)},
     \end{array}
     \label{RIS_spatial_correlation}
   \end{equation}
where $\text{sinc}(x)=\text{sin}(\pi x)/(\pi x)$ and $\lambda_c$ is the carrier wavelength. Moreover, the position vector follows $\textbf{u}_y=[0,\text{mod}(y-1,L_h)d_H,\lfloor(y-1)/L_h\rfloor d_V]^T$, $y\in\{m,n\}$\cite{9598875,10001167,9300189}, $L_h$ is the number of column elements and $L_v$ is the number of row elements at each RIS, with $L=L_h\times L_v$.

In this work, we assume that $\textbf{g}_{mj}^{\text{c}}[n]=\textbf{g}_{mj}^{\text{c}}$, $n=1,...,\tau_c$, remain constant within the resource block containing $\tau_c$ symbols, since only users are movable, while APs and RISs are stationary. The phase shift matrix of $j$-th RIS is $\boldsymbol{\Phi}_j[n]=\text{diag}(e^{i\boldsymbol{\phi}_{j,11}[n]},e^{i\boldsymbol{\phi}_{j,22}[n]},...,e^{i\boldsymbol{\phi}_{j,LL}[n]})\in\mathbb{C}^{L\times L}$, in which $e^{i\boldsymbol{\phi}_{j,ll}[n]}$ is the reflection coefficient of the $l$-th element in the $j$-th RIS. Moreover, the induced phase shift is $\boldsymbol{\phi}_{j,ll}[n]\in[0,2\pi)$\cite{10129196}. Since phase shift matrices are controllable, we assume these are pre-assigned and constant as $\boldsymbol{\Phi}_j[n]= \boldsymbol{\Phi}_j$, $n=1,...,\tau_c$ within the resource block. Besides, $\textbf{g}_{kj}^{\text{c}}[n]\in \mathbb{C}^{L\times 1}$, the channel from the $k$-th user to the $j$-th RIS, is given by
     \begin{equation}
     \begin{array}{c@{\quad}c}
\textbf{g}_{kj}^{\text{c}}[n]=\sqrt{\beta_{kj}}{\textbf{R}}_{kj}^{1/2}\textbf{v}_{kj}^\text{c}[n].
     \end{array}
     \label{channel_user_RIS}
   \end{equation}
Similar to \eqref{channel_RIS_AP}, $\beta_{kj}$ is the large-scale fading coefficient between the $k$-th user and $j$-th RIS. $\textbf{v}_{kj}^{\text{c}}[n]\in \mathbb{C}^{L\times 1}$, composed of i.i.d random variables following $\mathcal{CN}(0,1)$, represents the independent fast-fading channel. ${\textbf{R}}_{kj} =A_j{\textbf{R}_j} \in \mathbb{C}^{L\times L}$ is the one-sided correlation matrix at the $j$-th RIS\cite{9598875,9300189}. For the sake of exposition, the covariance matrix of $\textbf{g}_{mk}[n]$ can be obtained by
\begin{equation}
     \begin{array}{ll}
\displaystyle \mathbf{\Delta}_{mk} &= \mathbb{E}\{\textbf{g}_{mk}[n]\textbf{g}_{mk}[n]^H\}\vspace{2 pt}\\ & \displaystyle=\beta_{mk}\textbf{R}_{mk}+\sum\nolimits_{j=1}^J\beta_{mj}\beta_{kj}\textbf{R}_{mj,r}\text{tr}({\textbf{R}}_{mj,t}^{1/2}\boldsymbol{\Phi}_j\textbf{R}_{kj}\boldsymbol{\Phi}_j^H{\textbf{R}}_{mj,t}^{1/2})\vspace{2 pt}\\ & \displaystyle=\beta_{mk}\textbf{R}_{mk}+\sum\nolimits_{j=1}^J\beta_{mj}\beta_{kj}\textbf{R}_{mj,r}\text{tr}({\textbf{T}}_j),
     \end{array}
     \setlength{\belowdisplayskip}{3pt}
\label{RIS_spatial_correlation_element}
   \end{equation}
 with
 \begin{equation}
 \setlength{\abovedisplayskip}{3pt}
     \begin{array}{ll}
{\textbf{T}}_j \displaystyle={\textbf{R}}_{mj,t}^{1/2}\boldsymbol{\Phi}_j\textbf{R}_{kj}\boldsymbol{\Phi}_j^H{\textbf{R}}_{mj,t}^{1/2}\displaystyle=A_j^2{\textbf{R}}_j^{1/2}\boldsymbol{\Phi}_j\textbf{R}_{j}\boldsymbol{\Phi}_j^H{\textbf{R}}_j^{1/2}.
     \end{array}
     \label{T_j}
   \end{equation}
    \begin{equation}
     \begin{array}{ll}
\displaystyle \mathbf{\Delta}_{mk}^\text{d}=\beta_{mk}\textbf{R}_{mk},
     \end{array}
   \end{equation}
   \begin{equation}
     \begin{array}{ll}
\displaystyle \mathbf{\Delta}_{mk}^\text{c}=\sum\nolimits_{j=1}^J\beta_{mj}\beta_{kj}\textbf{R}_{mj,r}\text{tr}({\textbf{T}}_j).
     \end{array}
     \label{Delta_c}
   \end{equation}
   
%Given the pre-defined large-scale fading coefficients, the phase shift matrices and the spatial correlation matrices will determine $\mathbf{\Delta}_{mk}$.
%, which are all assumed to be known, as defined previously.

\subsection{Electromagnetic Interference Model} 
According to \cite{10167480,9598875,10134557}, uncontrollable external sources can generate the superposition of a continuum of incoming plane waves and result in non-negligible EMI that degrades system performance. The effect of EMI can be modelled by using the RIS correlation matrix under isotropic conditions\cite{9665300, 9598875,10167480}. Then, the EMI impinging on the $j$-th RIS can be given by
\begin{equation}
     \begin{array}{c@{\quad}c}
\displaystyle \textbf{n}_j\sim\mathcal{CN}(0,A_j\sigma_j^2\textbf{R}_{j}),
     \end{array}
     \label{EMI}
   \end{equation}
where $\sigma_j^2$ is the EMI power at the $j$-th RIS. In this work, we introduce a modified EMI power expression referring to \cite{10167480,9598875}, so that it is scalable and suitable for analysis
\begin{equation}
     \begin{array}{c@{\quad}c}
\displaystyle \sigma_j^2= \sqrt{\frac{p_up_d\sum\nolimits_{m=1}^M\beta_{mj}\sum\nolimits_{k=1}^K\beta_{kj}}{MK\varsigma_j^2}} ,
     \end{array}
     \label{EMI}
   \end{equation}
where $p_u$ and $p_d$ represent the respective uplink and downlink transmit power. The received signal power divided by the EMI power represents the ratio $\varsigma_j$ at the $j$-th RIS \cite{9598875,10167480}. For simplicity, we assume $\varsigma_j=\varsigma,~\forall j$.

\subsection{Channel Aging}
In general, user mobility generates relative movement between the users and APs/RISs, leading to a Doppler shift that changes the AP-user and RIS-user channels over time. This phenomenon is introduced as the channel aging effect\cite{9416909, 9875036}. As such, the channel coefficients within each symbol remain constant and change from symbol to symbol. Unlike the block-fading model, this channel-varying assumption is commonly used in works concerning the channel aging effect \cite{9416909,9838672,9875036}. Mathematically, the summation of the corresponding initial states $\textbf{g}_{mk}[0]$ and the evolution component can model the aggregate uplink channel $\textbf{g}_{mk}[n]$, which can be expressed as \cite{9416909,10468556,9765773,9406061}
\begin{equation}
     \displaystyle \textbf{g}_{mk}[n]=\rho_k[n]\textbf{g}_{mk}[0]+\bar{\rho}_k[n]\textbf{e}_{mk}[n],
     \label{uplink_channel_channel_aging}
   \end{equation}
where ${\rho}_k[n]=J_0(2\pi f_{D,k}T_sn)$ represents the temporal correlation coefficient combating the channel aging effect for the $k$-th user at time instant $n$ with $\bar{\rho}_k[n]=\sqrt{1-{\rho}_k^2[n]}$ \cite{9765773,7120183,9406061}. $J_0(\cdot)$ denotes the zero-order Bessel function of the first kind, $f_{D,k}=\frac{v_kf_c}{c}$ is the $k$-th user's Doppler shift, and $T_s$ is the time instant length. Moreover, $v_k$ is the $k$-th user velocity, $c=3\times 10^8~ m/s$ is the speed of light, and $f_c$ is the carrier frequency. We can see that a higher user velocity or delay can decrease $\rho_k[n]$. %This work focuses on channel aging and uses time correlation to combat it. 
Furthermore, the model in \eqref{uplink_channel_channel_aging} depends on its initial state at time instant 0, $\textbf{g}_{mk}[0]$, to match with the statistics of Jakes' model \cite{9875036,9905943} \cite{8388873}. 

According to \eqref{aggregate_uplink_channel}, the channel coefficients $\textbf{g}_{mk}^{\text{d}}[n]$ and $\textbf{g}_{mkj}^{\text{c}}[n]$ can be expressed by their respective initial states $\textbf{g}_{mk}^{\text{d}}[0]$, $\textbf{g}_{mkj}^{\text{c}}[0]$ and the perturbation terms\cite{9905943}. Then, we can obtain $\textbf{g}_{mk}[n]$ at time instant $n$ as

\begin{equation}
\setlength{\abovedisplayskip}{3pt}
\begin{array}{ll}
     \displaystyle \textbf{g}_{mk}[n]%\rho_k[n]\textbf{g}_{mk}[0]+\bar{\rho}_k[n]\textbf{e}_{mk}[n]\\
   \displaystyle =\displaystyle  \rho_k[n]\textbf{g}_{mk}^{\text{d}}[0]+
     \bar{\rho}_k[n]\textbf{e}_{mk}^{\text{d}}[n]\vspace{2 pt}\\
     ~~~~~~~+     \displaystyle \sum\nolimits_{j=1}^{J}\rho_k[n]\textbf{g}_{mkj}^{\text{c}}[0]+\sum\nolimits_{j=1}^{J}\bar \rho_k[n]\textbf{e}_{mkj}^{\text{c}}[n],
\end{array}\label{uplink_channel_channel_aging_decomposion}
\setlength{\belowdisplayskip}{3pt}
   \end{equation}
where $\textbf{e}_{mk}^{\text{d}}[n]$ and  $\textbf{e}_{mkj}^{\text{c}}[n]$ are the independent AP-user and RIS-AP-user innovation component
with $\textbf{e}_{mk}^{\text{d}}[n]\sim\mathcal{CN}(\textbf{0},\beta_{mk}\textbf{R}_{mk})$ and $\textbf{e}_{mkj}^{\text{c}}[n]=\textbf{g}_{mj}^{\text{c}}{\theta}_j[n]\textbf{e}_{kj}^{\text{c}}[n]\sim\mathcal{CN}(\textbf{0},\beta_{mj}\beta_{kj}\textbf{R}_{mj,r}\text{tr}({\textbf{T}}_j))$, where $\textbf{e}_{kj}^{\text{c}}[n]\sim\mathcal{CN}(\textbf{0},\beta_{kj}A_j\textbf{R}_{j})$, respectively.

\section{Novel Two-Phase Channel Estimation}

Usually, users first send the pilots to APs, which adopt the minimum mean square error (MMSE) estimate to acquire the uplink channel state information (CSI)\cite{10225319}. Then, the obtained uplink CSI will be applied to introduce the channel distribution of other time instants with the temporal correlation coefficients across different time instants \cite{9905943,7473866,ref}.
To improve the channel estimation accuracy, we extend our novel two-phase channel estimation scheme \cite{10621117} to enhance channel estimation accuracy and mitigate the effect of channel aging and EMI. Compared with the conventional MMSE estimation scheme, which simultaneously estimates the aggregate channels (both direct and RIS-assisted channels) and might cause poor channel estimation accuracy \cite{10225319,9322151,9905943,ref}, our proposed scheme divides the channel estimation phase into two sub-phases, each taking $\tau_p$ samples. The two sub-phases estimate the direct and RIS-assisted channels in sequence, and each sub-phase involves fractional power control-aided pilot assignment for channel estimation improvement. %This channel estimation process is described next.

\subsection{Large-Scale Fading Coefficient-based Pilot Assignment}
%\subsection{Fractional Power Control-aided Pilot Assignment}
%\subsubsection{Fractional Power Control} In the uplink channel estimation phase, we utilize the user's large-scale fading coefficients to introduce an appropriate fractional power control to counteract the near-far effects and improve the estimation accuracy\cite{8968623,9322468}. As such, the pilot power of the $k$-th user at the $\text{x}$-the sub-phase, $p_k^{\text{x}}$, can be obtained by
%\begin{equation}
%\begin{array}{ll}
  %   \displaystyle p_k^{\text{x}}=\left(\frac{\text{min}_{\forall k}\sum\nolimits_{m=1}^M\text{tr}(\mathbf{\Delta}_{mk}^{\text{x}})}{\sum\nolimits_{m=1}^M\text{tr}(\mathbf{\Delta}_{mk'}^{\text{x}})}\right)^\alpha p_{p},
  %   \end{array}
  %   \label{MMSE_pilot_power}
  % \end{equation}
%where $\text{x}=\text{d},\text{c}$ represents the direct channel estimation and RIS-assisted channel estimation sub-phase, respectively. Moreover, $p_p$ is the pilot power per user without power control and $0\leq \alpha \leq 1$ is the fractional power control parameter \cite{10264149}.
%\subsubsection{Pilot Assignment}
In the uplink channel estimation phase, we utilize the user's large-scale fading coefficients to introduce an appropriate pilot assignment to reduce the pilot contamination and guarantee a certain
QoS for all users\cite{10264149}. First, for the $\text{x}$-th sub-phase, the $m_k^\text{max,x}$-th AP with the largest $\mathbf{\Delta}_{mk}^{\text{x}}$ can be chosen as the prime AP of the $k$-th user,
\begin{equation}
\begin{array}{ll}
     \displaystyle m_k^\text{max,x}=\text{arg}~\text{max}_{m\in(1,...,M)}~\text{tr}(\mathbf{\Delta}_{mk}^{\text{x}}),
     \end{array}
   \end{equation}
where $\text{x}=\text{d},\text{c}$ denotes the ongoing sub-phase and subsequently, the $m_k^\text{max,x}$-th AP determines the pilot sequence allocation of $\boldsymbol{\varphi}_k$ to the $k$-th user during the relevant sub-phase. To mitigate pilot contamination, the pilot sequence with index $t_k^\text{x}=k$ is allocated to the $k$-th user when $k<\tau_p$, and when $k>\tau_p$, $t_k$ follows
\begin{equation}
\begin{array}{ll}
        \displaystyle   t_k^\text{x}=\text{arg}~\text{min}_{t\in\{1,2,...,\tau_p\}}~
 \sum\nolimits_{i\in\mathcal{S}_t} \text{tr}(\mathbf{\Delta}_{{m_k^\text{max,x}}i}^{\text{x}}) ,
     \end{array}
     \label{pilot_assignment}
   \end{equation}
where $\mathcal{S}_t$ is the set of users assigned with pilot sequence $t$, and $\mathcal{S}_t$ will be updated when each $k$-th user ($k>\tau_p$) is assigned with its own pilot sequence $t_k^\text{x}$ based on \eqref{pilot_assignment}. With the help of the pilot assignment, we can determine $\mathcal{P}_k^{\text{x}}$ (${\text{x}}={\text{d}},{\text{c}}$, $k=1,...,K$), as the set of users sharing the same pilot sequence including the $k$-th user itself in the ongoing $\text{x}$-th sub-phase.

\subsection{Uplink Channel Estimation}

 We define the pilot sequence allocated to the $k$-th user as $\boldsymbol{\varphi}_k \in \mathbb{C}^{\tau_p\times 1}$ with $\boldsymbol{\varphi}_k^{H}\boldsymbol{\varphi}_k=1,\forall k$, where $\tau_p$ is the pilot length. Usually, the coherence interval length $\tau_c$ is much larger than $\tau_p$ to guarantee the transmission efficiency, implying that $K>\tau_p$ \cite{9665300,10167480}. Different users will share the same pilot sequence in this case, leading to pilot contamination \cite{5898372}. According to the pilot assignment procedure, $\mathcal{P}_k^{\text{x}}$ is the set of users sharing the same pilot in the ongoing sub-phase satisfying $\boldsymbol{\varphi}_k^{H}\boldsymbol{\varphi}_{k'}=1, \forall k' \in \mathcal{P}_k^{\text{x}}$, inevitably.

\subsubsection{Direct channel estimation sub-phase} All RIS elements are OFF within the current sub-phase. Then, all users transmit their pilots to the APs to estimate the direct channel. $\tau_p$ pre-defined mutually orthogonal pilot sequences are used in this sub-phase. Referring to the channel estimation model experiencing channel aging in \cite{9416909}, the pilot sequence $t$ is sent only at time instant $t$ within the sub-phase to maintain the orthogonality between the pilots with channel aging. $t_k ^{\text{d}}\in{1,...,\tau_p}$ is the time instant index assigned to the $k$-th user, and the other users in the pre-defined set $\mathcal{P}_k^\text{d}$ share the same pilot transmission time instant, namely, $\mathcal{P}_k^\text{d}=\{k'~:~t_{k'}^{\text{d}}=t_k^{\text{d}}\}\subset\{1,...,K\}$.
The pilot signal, $\textbf{Y}_{m,p}^{\text{d}} [t_k]\in \mathbb{C}^{N\times\tau_p}$, received at the $m$-th AP at the $t_k^{\text{d}}$-th time instant is formulated as
\begin{equation}
\begin{array}{ll}
     \displaystyle \textbf{Y}_{m,p}^{\text{d}}[t_k^\text{d}]=\sum\nolimits_{k'=1}^{K}\sqrt{p_p}\textbf{g}_{mk'}^{\text{d}}[t_{k'}^\text{d}]\boldsymbol{\varphi}_{k'}^{H}+\textbf{W}^{\text{d}}_{m,p}[t_k^\text{d}],
\end{array}\label{received_pilot_signal_1st_phase}
   \end{equation}
where $p_p$ is the pilot power for each user. $\textbf{W}^{\text{d}}_{m,p}[t_k^\text{d}]$ is the additive white Gaussian noise (AWGN) at the $m$-th AP during this sub-phase with $v$-th column satisfying
$\Big{[}\textbf{W}^{\text{d}}_{m,p}[t_k^\text{d}]\Big{]}_v\sim \mathcal{CN}(\textbf{0},\sigma^2\textbf{I}_{N})$, where $\sigma^2$ is the noise power. We focus on the channel estimates at $\lambda$-th time instant in the following, where $\lambda=2\tau_p+1$. Then, these estimates will be treated as the initial states to generate the other time instants' channel estimates. We can obtain the effective channel at the $t_{k'}^{\text{d}}$-th time instant based on the channel at the $\lambda$-th time instant as
\begin{equation}
\begin{array}{ll}
\displaystyle \textbf{g}_{mk'}^\text{d}[t_{k'}^\text{d}]=\rho_{k'}[\lambda-t_{k'}^\text{d}]\textbf{g}_{mk'}^\text{d}[\lambda]+\bar{\rho}_{k'}[\lambda-t_{k'}^\text{d}]\textbf{e}_{mk'}^\text{d}[t_{k'}^\text{d}].
\end{array}
\label{eq_19}
   \end{equation}
By projecting $\textbf{Y}_{m,p}^{\text{d}}[t_k^\text{d}]$ on $\boldsymbol{\varphi}_k$, we can obtain
\begin{equation}
\begin{array}{ll}
     \displaystyle \textbf{y}_{mk,p}^{\text{d}}[t_k^\text{d}]&   \displaystyle =\sum\nolimits_{{k'}=1}^{K}\sqrt{p_p}\textbf{g}_{m{k'}}^{\text{d}}[t_{k'}^\text{d}]\boldsymbol{\varphi}_{k'}^{H}\boldsymbol{\varphi}_k+\textbf{W}^{\text{d}}_{m,p}[t_k^\text{d}]\boldsymbol{\varphi}_k\vspace{2 pt}\\&\displaystyle=\sum\nolimits_{{k'}\in\mathcal{P}_k^\text{d}}\sqrt{p_p}\textbf{g}_{m{k'}}^{\text{d}}[t_{k'}^\text{d}]+\textbf{W}^{\text{d}}_{m,p}[t_{k}^\text{d}]\boldsymbol{\varphi}_k.
\end{array}\label{projected_pilot_signal_1st_phase}
   \end{equation}
   Given \eqref{projected_pilot_signal_1st_phase}, the MMSE estimation of $\textbf{g}_{mk}^{\text{d}}[\lambda]$ is given by
   \begin{equation}
\begin{array}{ll}
     \displaystyle \hat{\textbf{g}}_{mk}^{\text{d}}[\lambda]=\textbf{R}_{mk}^\text{d}(\mathbf{\Psi}_{mk}^\text{d})^{-1}\textbf{y}_{mk,p}^{\text{d}}[t_k^\text{d}],
\end{array}\label{LMMSE_direct_channel_estimation}
   \end{equation}
   where 
   \begin{equation}
   \setlength{\abovedisplayskip}{5pt}
\begin{array}{ll}
     \displaystyle 
     \textbf{R}_{mk}^\text{d}
=\sqrt{p_p}\rho_k[\lambda-t_k^\text{d}]\beta_{mk}\textbf{R}_{mk},
\end{array}\label{D_est1}
   \end{equation}
      \begin{equation}
      \setlength{\abovedisplayskip}{5pt}
\begin{array}{ll}
     \displaystyle 
\mathbf{\Psi}_{mk}^\text{d}=\sum\nolimits_{k'\in\mathcal{P}_k^{\text{d}}}{p_p}\beta_{mk'}^{\text{d}}{\textbf{R}}_{mk'}+\sigma^2\textbf{I}_{\text{N}}.
\end{array}\label{D_est2}
   \end{equation}
    
\subsubsection{RIS-assisted channel estimation sub-phase} All RIS elements are ON within this sub-phase. Similarly, the pre-defined $\tau_p$ pilot sequences are applied, and the pilot sequence $t$ is sent only at time instant $t+\tau_p$ during the current sub-phase\cite{9416909}. $t_k^\text{c} \in\{\tau_p,...,\tau_p\}$ is the pilot sequence assigned to users utilizing the same time instant as the $k$-th user, follows $\mathcal{P}_k^\text{c}=\{k'~:~t_{k'}^\text{c}=t_k^\text{c}\}\subset\{1,...,K\}$. For simplicity of notation, $t_k^\text{c}=\tau_p+t_k^\text{c} \in\{\tau_p+1,...,2\tau_p\}$ represents the relevant time instant index during the current sub-phase.
As such, $\textbf{Y}_{m,p}[t_k]\in \mathbb{C}^{N\times\tau_p}$,
the received pilot signal at the $m$-th AP, can be modelled as
\begin{equation}
\begin{array}{ll}
     \displaystyle \textbf{Y}_{m,p}[t_k^\text{c}]&      \displaystyle =\sum\nolimits_{k'=1}^{K}\sqrt{p_p}\Big{(}\textbf{g}_{mk'}^{\text{d}}[t_{k'}^\text{c}]+\sum\nolimits_{j=1}^J\textbf{g}_{mk'j}^{\text{c}}[t_{k'}^\text{c}]\Big{)}\boldsymbol{\varphi}_{k'}^{H}\vspace{2 pt}\\&     \displaystyle+\sum\nolimits_{j=1}^J\textbf{g}_{mj}^{\text{c}}\boldsymbol{\Phi}_{j}\textbf{N}_j[t_k^\text{c}]+\textbf{W}_{m,p}^{\text{c}}[t_k^\text{c}],
\end{array}\label{received_pilot_signal_jth_phase}
   \end{equation}
where $\textbf{W}_{mj,p}^{\text{c}}[t_k^\text{c}]$ is the AWGN matrix in this sub-phase with $\Big{[}\textbf{W}_{m,p}^{\text{c}}[t_k^\text{c}]\Big{]}_v\sim \mathcal{CN}(\textbf{0},\sigma^2\textbf{I}_{N})$. Moreover, $\textbf{N}_j[t_k^\text{c}]\in\mathbb{C}^{L\times\tau_p}$ is the EMI of the $j$-th RIS with $v$-th column satisfying $\Big{[}\textbf{N}_j[t_k^\text{c}]\Big{]}_v\sim \mathcal{CN}(0,A_j\sigma_j^2\textbf{R}_{j})$.
Since the direct link has already been estimated in the first sub-phase and $\hat{\textbf{g}}_{mk}^{\text{d}}[\lambda]$ is known at the $m$-th AP. We consider the perfect subtraction of $\textbf{Y}_{m,p}^{\text{d}}[t_k^\text{c}]$ from $\textbf{Y}_{m,p}[t_k^\text{c}]$ in the this work, the imperfect subtraction will be left for future work. Then, we can obtain
\begin{equation}
\begin{array}{ll}
     \displaystyle \textbf{Y}_{m,p}^{\text{c}}[t_k^\text{c}]& \displaystyle =\sum\nolimits_{k'=1}^{K}\sum\nolimits_{j=1}^J\sqrt{p_p}\textbf{g}_{mk'j}^{\text{c}}[t_{k'}^\text{c}]\boldsymbol{\varphi}_{k'}^{H}\vspace{2 pt}\\& \displaystyle +\sum\nolimits_{j=1}^J\textbf{g}_{mj}^{\text{c}}\boldsymbol{\Phi}_{j}\textbf{N}_j[t_k^\text{c}]+\textbf{W}_{mj,p}^{\text{c}}[t_k^\text{c}],
\end{array}\label{received_pilot_signal_j-th_phase}
   \end{equation}
   similar to \eqref{eq_19}, the effective channel at the $t_{k'}^\text{c}$-th time instant based on the channel at the $\lambda$-th time instant is expressed as
\begin{equation}
\begin{array}{ll}
\displaystyle \textbf{g}_{mk'}^\text{c}[t_{k'}^\text{c}]&\displaystyle=\rho_{k'}[\lambda-t_{k'}^\text{c}]\sum\nolimits_{j=1}^J\textbf{g}_{mk'j}^\text{c}[\lambda]\vspace{2 pt}\\&\displaystyle+\bar{\rho}_{k'}[\lambda-t_{k'}^\text{c}]\sum\nolimits_{j=1}^J\textbf{e}_{mk'j}^\text{c}[t_{k'}^\text{c}].
\end{array}
   \end{equation}
The projection of $\textbf{Y}_{m,p}^{\text{c}}[t_k^\text{c}]$ on $\boldsymbol{\varphi}_k$ can be formulated as
\begin{equation}
\begin{array}{ll}
     \displaystyle \textbf{y}_{mk,p}^{\text{c}}[t_k^\text{c}]%& \displaystyle =\sum\limits_{{k'}=1}^{K}\sum\limits_{j=1}^J\sqrt{p^\text{c}_{k'}}\textbf{g}_{m{k'}j}^{\text{c}}[t_{k'}]\boldsymbol{\varphi}_{k'}^{H}\boldsymbol{\varphi}_k\vspace{2 pt}\\&\displaystyle+\sum\limits_{k'=1}^{K}\sqrt{p_{k'}^c}\tilde{\textbf{g}}_{mk'}^{\text{d}}[t_{k'}]\boldsymbol{\varphi}_{k'}^{H}\boldsymbol{\varphi}_k\\& \displaystyle +\sum\limits_{j=1}^J\textbf{g}_{mj}^{\text{c}}\boldsymbol{\Phi}_{j}\textbf{N}_j[t_k]\boldsymbol{\varphi}_k+\textbf{W}_{m,p}^{\text{c}}[t_k]\boldsymbol{\varphi}_k\vspace{2 pt}
     & \displaystyle =\sum\nolimits_{{k'}\in\mathcal{P}_{k}^{\text{c}}}\sum\nolimits_{j=1}^J\sqrt{p_p}\textbf{g}_{m{k'}j}^{\text{c}}[t_{k'}^\text{c}]\\& \displaystyle +\sum\nolimits_{j=1}^J\textbf{g}_{mj}^{\text{c}}\boldsymbol{\Phi}_{j}\textbf{N}_j[t_k^\text{c}]\boldsymbol{\varphi}_k+\textbf{W}_{m,p}^{\text{c}}[t_k]\boldsymbol{\varphi}_k.
\end{array}\label{projected_pilot_signal_j-th_phase}
   \end{equation}
Given \eqref{projected_pilot_signal_j-th_phase}, the MMSE of $\textbf{g}_{mk}^{\text{c}}[\lambda]=\sum\nolimits_{j=1}^{J}\textbf{g}_{mkj}^{\text{c}}[\lambda]$ is defined as
\begin{equation}
\begin{array}{ll}
     \displaystyle \hat{\textbf{g}}_{mk}^{\text{c}}[\lambda]=\textbf{R}_{mk}^\text{c}(\mathbf{\Psi}_{mk}^\text{c})^{-1}\textbf{y}_{mk,p}^{\text{c}}[t_k^\text{c}],
\end{array}\label{LMMSE_RIS_channel_estimation}
   \end{equation}
   where
   \begin{equation}
   \setlength{\abovedisplayskip}{5pt}
\begin{array}{ll}
\displaystyle\textbf{R}_{mk}^\text{c}=\sqrt{p_p}\rho_k[\lambda-t_k^\text{c}]\sum\nolimits_{j=1}^J\beta_{mj}\beta_{kj}\text{tr}({\textbf{T}}_j)\textbf{R}_{mj,r},
\end{array}\label{R_est1}
\setlength{\belowdisplayskip}{5pt}
   \end{equation}
 \begin{equation}
 \setlength{\abovedisplayskip}{5pt}
\begin{array}{ll}
\displaystyle
\mathbf{\Psi}_{mk}^\text{c}\displaystyle=\sum\nolimits_{k'\in\mathcal{P}_k^{\text{c}}}{p_p}\boldsymbol{\Delta}_{mk'}^\text{c}\displaystyle+\sum\nolimits_{j=1}^J\beta_{mj}\sigma_j^2\text{tr}({\textbf{T}}_j)\textbf{R}_{mj,r}+\sigma^2\textbf{I}_N.
\end{array}
\label{Psi_c}
   \end{equation}
   
Based on the above, the sum of the direct channel estimate and cascaded RIS-assisted channel estimates introduces the MMSE channel estimate $\hat{\textbf{g}}_{mk}[\lambda]$, which can be formulated as %of the aggregated channel $\textbf{g}_{mk}[\lambda]$ 
\begin{equation}
\begin{array}{ll}
     \displaystyle \hat{\textbf{g}}_{mk}[\lambda]=\hat{\textbf{g}}_{mk}^{\text{d}}[\lambda]+\hat{\textbf{g}}_{mk}^{\text{c}}[\lambda].
     \end{array}
     \label{aggregate_uplink_channel_estimation}
   \end{equation}
   
  The channel estimate $\hat{\textbf{g}}_{mk}[\lambda]$ and the channel estimation error $\tilde{\textbf{g}}_{mk}[\lambda]={\textbf{g}}_{mk}[\lambda]-\hat{\textbf{g}}_{mk}[\lambda]$ are distributed as $\mathcal{CN}\left(\textbf{0},\textbf{Q}_{mk}^\text{d}+\textbf{Q}_{mk}^\text{c}\right)$ and $\mathcal{CN}\left(\textbf{0},\mathbf{\Delta}_{mk}-\textbf{Q}_{mk}^\text{d}-\textbf{Q}_{mk}^\text{c}\right)$, respectively, with $\textbf{Q}_{mk}^\text{x}=\textbf{R}_{mk}^\text{x}(\textbf{R}_{mk}^\text{x}(\mathbf{\Psi}_{mk}^\text{x})^{-1})^H$, $\text{x}\in\{\text{d},\text{c}\}$.
   %\begin{equation}
%\begin{array}{ll}
   %  \displaystyle \textbf{Q}_{mk}^\text{d}=\textbf{R}_{mk}^\text{d}(\textbf{R}_{mk}^\text{d}\mathbf{\Psi}_{mk}^\text{d})^\text{H},
   %  \end{array}
   %  \label{Num_direct}
  % \end{equation}
     % \begin{equation}
%\begin{array}{ll}
    % \displaystyle \textbf{Q}_{mk}^\text{c}=\textbf{R}_{mk}^\text{c}(\textbf{R}_{mk}^\text{c}\mathbf{\Psi}_{mk}^\text{c})^\text{H}.
     %\end{array}
 %  \end{equation}
   
   Moreover, to study the channel estimation accuracy, the normalised mean square error (NMSE) of the channel estimation is modelled as \cite{10225319,8388873}
 \begin{equation}
\begin{array}{ll}
     \displaystyle \text{NMSE}=\displaystyle\frac{\displaystyle \sum\nolimits_{m,k}\mathbb{E}\{||\tilde{\textbf{g}}_{mk}[\lambda]||^2\}}{\displaystyle \sum\nolimits_{m,k}\mathbb{E}\{||{\textbf{g}}_{mk}[\lambda]||^2\}}=\frac{\displaystyle \sum\nolimits_{m,k}\text{tr}(\mathbf{\Delta}_{mk}-\textbf{Q}_{mk}^\text{d}-\textbf{Q}_{mk}^\text{c})}{\displaystyle \sum\nolimits_{m,k}\text{tr}(\mathbf{\Delta}_{mk})}.
     \end{array}
     \label{NMSE_expression}
   \end{equation}
   
In this case, $\textbf{g}_{mk}[n]$, taking the joint channel aging and channel estimation error into consideration, can be formulated as \cite{9905943,8388873}
\begin{equation}
\begin{array}{ll}
     \displaystyle \textbf{g}_{mk}[n]
   \displaystyle =\displaystyle  \rho_k[n-\lambda]\Big{(}\hat{\textbf{g}}_{mk}^{\text{d}}[\lambda]+\tilde{\textbf{g}}_{mk}^{\text{d}}[\lambda]\Big{)}+
     \bar{\rho}_k[n-\lambda]\textbf{e}_{mk}^{\text{d}}[n]\\
     ~~~~~~~~+     \displaystyle \rho_k[n-\lambda]\Big{(}\hat{\textbf{g}}_{mk}^{\text{c}}[\lambda]+\tilde{\textbf{g}}_{mk}^{\text{c}}[\lambda]\Big{)}+\sum\nolimits_{j=1}^{J}\bar \rho_k[n-\lambda]\textbf{e}_{mkj}^{\text{c}}[n].
\end{array}
\label{channel_estimate}
   \end{equation}

\section{Downlink SE Analysis and Sum SE Maximization}
This section analyzes the downlink data transmission affected by channel aging and EMI. After uplink channel estimation, $\tau_c-2\tau_p$ time instants are adopted for the downlink transmission. We derive novel downlink SE closed-from expressions and utilize downlink fractional power control to evaluate system performance. We then find the RIS coefficient matrix design to maximize the sum downlink SE.

\subsection{Downlink Data Transmission}

The downlink data transmission from all APs to all users is composed of a broadcast channel utilizing the beamforming vector $\textbf{f}_{mk}[n]\in\mathbb{C}^{N\times 1}$ at the $m$-th AP\cite{9875036}.
We treat the uplink channel transpose as the downlink channel based on the channel reciprocity characteristic of TDD operation.\cite{9765773,8388873}. As such, the transmit signal at time instant $n$ from the $m$-th AP is formulated as
\begin{equation}
\begin{array}{ll}
     \displaystyle \textbf{x}_m[n]=\sqrt{p_d}\sum\nolimits_{k=1}^K \textbf{f}_{mk}[n]\sqrt{\eta_{mk}}q_k[n],
\end{array}\label{transmitted_signal_m_AP}
   \end{equation}
where $p_d$ denotes the downlink transmit power at APs, $\eta_{mk}$ is downlink power control coefficients with $\mathbb{E}\big{\{}|\textbf{x}_m[n]|^2\big{\}}\leq p_d$. $q_k[n]\sim \mathcal{CN}(0,1)$ is the signal sent to the $k$-th user, same for all APs. Conjugate beamforming at APs is selected, namely, $\textbf{f}_{mk}[n]=\hat{\textbf{g}}_{mk}^{\text{*}}[\lambda]$ for $\lambda\leq n\leq \tau_c$.
Then, \eqref{downlink_received_signal_k_user} describes the $k$-th user's received signal at the top of the next page. $\textbf{n}_j[n]\sim\mathcal{CN}(0,A_j\sigma_j^2\textbf{R}_{j})$ is the EMI of the $j$-th RIS, $w_k[n]\sim \mathcal{CN}(0,\sigma^2)$ is the AWGN at the $k$-th user at time instant $n$.
\begin{figure*}[!t]
\begin{equation}
\begin{array}{ll}
     \displaystyle r_k[n]&   \displaystyle =\sum\limits_{m=1}^{M}\textbf{g}_{mk}^{T}[n]\textbf{x}_m[n]+\sum\limits_{j=1}^{J}(\textbf{g}_{kj}^{\text{c}}[n])^T\boldsymbol{\Phi}_j^T\textbf{n}_j[n]+w_k[n]\\&   \displaystyle =\sum\limits_{m=1}^{M}\sum\limits_{k'=1}^K\textbf{g}_{mk}^{T}[n]\sqrt{\rho_d} \hat{\textbf{g}}_{mk'}^*[\lambda]\sqrt{\eta_{mk'}}q_{k'}[n]+\sum\limits_{j=1}^{J}(\textbf{g}_{kj}^{\text{c}}[n])^T\boldsymbol{\Phi}_j^T\textbf{n}_j[n]+w_{k}[n]\\&=\underbrace {\sqrt{p_d}\rho_k[n-\lambda]\sum\limits_{m=1}^{M}\mathbb{E}\Big{\{}\textbf{g}_{mk}^{T}[\lambda]\hat{\textbf{g}}_{mk}^*[\lambda]\Big{\}}\sqrt{\eta_{mk}}q_k[n]}_{\text{DS}_k[n]}\displaystyle+\underbrace {\sqrt{p_d}\rho_k[n-\lambda]\sum\limits_{m=1}^{M}\Big{(}\textbf{g}_{mk}^{T}[\lambda]\hat{\textbf{g}}_{mk}^*[\lambda]-\mathbb{E}\Big{\{}\textbf{g}_{mk}^{T}[\lambda]\hat{\textbf{g}}_{mk}^*[\lambda]\Big{\}}\Big{)}\sqrt{\eta_{mk}}q_k[n]}_{\text{BU}_k[n]}\\&\displaystyle+\underbrace {\sqrt{p_d}\bar{\rho}_k[n-\lambda]\sum\limits_{m=1}^{M}\textbf{e}_{mk}^{T}[n]\hat{\textbf{g}}_{mk}^*[\lambda]\sqrt{\eta_{mk}}q_k[n]}_{\text{CA}_k[n]}\displaystyle+\displaystyle\sum\limits_{k'\neq k}^K\underbrace {\sqrt{\rho_d} \sum\limits_{m=1}^{M}\textbf{g}_{mk}^{T}[n]\hat{\textbf{g}}_{mk'}^*[\lambda]\sqrt{\eta_{mk'}}q_{k'}[n]}_{\text{UI}_{kk'}[n]}+\underbrace{\sum\limits_{j=1}^{J}(\boldsymbol{\Phi}_j\textbf{g}_{kj}^{\text{c}}[n])^T\textbf{n}_j[n]}_{\text{EMI}_k[n]}+\underbrace {w_{k}[n]}_{\text{NS}_k[n]},
\end{array}\label{downlink_received_signal_k_user}
\vspace{-5 pt}
   \end{equation}
   \hrulefill
%\vspace{-8 pt}
\end{figure*}

Motivated by \cite{8968623,9875036}, we propose a fractional power control scheme including statistical channel information and generate the fractional power control coefficients as
\begin{equation}
\begin{array}{ll}
     \displaystyle \eta_{mk}=\left(\sum\nolimits_{k'=1}^K{\text{tr}(\textbf{Q}_{mk'}^\text{d}+\textbf{Q}_{mk'}^\text{c})}\right)^{-1},~\forall k,~\forall m.
\end{array}
   \end{equation}
\subsection{Performance Analysis and Closed-form SE Derivations}
According to \eqref{downlink_received_signal_k_user}, the downlink achievable SE of the $k$-th user is lower bounded by \cite{10225319,9416909}
\begin{equation}
\begin{array}{ll}
     \displaystyle \text{SE}_k=\frac{1}{\tau_c}\sum\nolimits_{n=\lambda}^{\tau_c}\text{log}_2\Big{(} 1+\text{SINR}_k[n]\Big{)},
\end{array}\label{downlink_SE_description}
   \end{equation}
where $\text{SINR}_k[n]$ is the effective signal-to-interference-plus-noise ratio (SINR) at time instant $n$ can be expressed as,
%\begin{figure*}[t!]
\begin{equation}
\begin{array}{ll}
\displaystyle \text{SINR}_k[n]
\displaystyle=\frac{S_k[n]}{I_k[n]},
\end{array}\label{downlink_SINR}
   \end{equation}
%\hrulefill
 %  \end{figure*}
 where $S_k[n]$ and $I_k[n]$ are expressed as \eqref{S_k} and \eqref{I_k} at the top of this page.
 \begin{figure*}[t!]
 \vspace{-6 pt}
\begin{equation}
\begin{array}{ll}
    \displaystyle  S_k[n]={p_d}\rho_k^2[n-\lambda]\Big{|}\sum\nolimits_{m=1}^{M}\sqrt{\eta_{mk}}\mathbb{E}\bigg{\{}\textbf{g}_{mk}^{T}[\lambda]\hat{\textbf{g}}_{mk}^*[\lambda]\bigg{\}}\Big{|}^2=p_d\rho_k^2[n-\lambda]\Big{|}\sum\nolimits_{m=1}^{M}\sqrt{\eta_{mk}}\text{tr}\left(\textbf{Q}_{mk}^\text{d}+\textbf{Q}_{mk}^\text{c}\right)\Big{|}^2,
     \end{array}\label{S_k}\vspace{-3 pt}
   \end{equation}
\hrulefill
   \end{figure*}
    \begin{figure*}[t!]
    \vspace{-6 pt}
\begin{equation}
\begin{array}{ll}
   \displaystyle I_k[n]={p_d}\sum\nolimits_{k'=1}^K\mathbb{E}\bigg{\{}\Big{|}\sum\nolimits_{m=1}^{M}\sqrt{\eta_{mk'}}\textbf{g}_{mk}^{T}[n]\hat{\textbf{g}}_{mk'}^*[\lambda]\Big{|}^2\bigg{\}}-S_k[n]+\mathbb{E}\Big{\{}\Big{|}\sum\nolimits_{j=1}^{J}(\boldsymbol{\Phi}_j\textbf{g}_{kj}^{\text{c}}[n])^T\textbf{n}_j[n]\Big{|}^2\Big{\}}+\sigma^2\\~~~~~~=\displaystyle{p_d}\sum\nolimits_{k'=1}^K\mathbb{E}\bigg{\{}\Big{|}\sum\nolimits_{m=1}^{M}\sqrt{\eta_{mk'}}\textbf{g}_{mk}^{T}[n]\hat{\textbf{g}}_{mk'}^*[\lambda]\Big{|}^2\bigg{\}}-S_k[n]+\sum\nolimits_{j=1}^{J}\beta_{kj}\sigma_j^2\text{tr}(\textbf{T}_j)+\sigma^2.
   \end{array}\label{I_k}
    \vspace{-3 pt}
   \end{equation}
\hrulefill
   \end{figure*}
\eqref{DL_downlink} shows the inter-user interference power summation at the top of the next page, where $\textbf{Z}_{mk,d}=\textbf{R}_{mk}^\text{d}(\mathbf{\Psi}_{mk}^\text{d})^{-1}, ~\textbf{Z}_{mk,c}=\textbf{R}_{mk}^\text{c}(\mathbf{\Psi}_{mk}^\text{c})^{-1},~\forall m,~\forall k$.
\begin{figure*}[t!]
\begin{equation}
\begin{array}{ll}
\vspace{2 pt}
&\displaystyle\sum\nolimits_{k'=1}^K\mathbb{E}\bigg{\{}\Big{|}\sum\nolimits_{m=1}^{M}\sqrt{\eta_{mk'}}\textbf{g}_{mk}^{T}[n]\hat{\textbf{g}}_{mk'}^*[\lambda]\Big{|}^2\bigg{\}} \vspace{2 pt}\\&\displaystyle=\sum\nolimits_{k'=1}^K\sum\nolimits_{m=1}^{M}{\eta_{mk'}}\text{tr}\left((\textbf{Q}_{mk'}^\text{d}+\textbf{Q}_{mk'}^\text{c})\mathbf{\Delta}_{mk}\right)\vspace{2 pt}\\&\displaystyle+\sum\nolimits_{k'\in\mathcal{P}_k^\text{c}}\sum\nolimits_{m=1}^{M}{\eta_{mk'}}\rho_k^2[n-\lambda]p_p\rho_k^2[\lambda-t_k^\text{c}]\sum\nolimits_{j=1}^J\beta_{mj}^2\beta_{kj}^2\text{tr}(\textbf{T}_j^2)\text{tr}(\textbf{R}_{mj,r}\textbf{Z}_{mk',c}\textbf{R}_{mj,r}\textbf{Z}^H_{mk',c})\vspace{2 pt}
\\& \displaystyle+\sum\nolimits_{k'=1}^K\sum\nolimits_{m=1}^{M}\sum\nolimits_{m'=1}^{M}\sqrt{\eta_{mk'}\eta_{m'k'}}\sum\nolimits_{j=1}^J\left(\sum\nolimits_{k''\in\mathcal{P}_{k'}^\text{c}}p_p\beta_{k''j}+\sigma^2_j\right)\beta_{mj}\beta_{m'j}\beta_{kj}\text{tr}(\textbf{T}_j^2)\text{tr}(\textbf{R}_{mj,r}\textbf{Z}_{mk',c}\textbf{R}_{m'j,r}\textbf{Z}^H_{m'k',c})\vspace{2 pt}
\\& \displaystyle+\sum\nolimits_{k'\in\mathcal{P}_k^\text{d}\cap\mathcal{P}_k^\text{c}}\sum\nolimits_{m=1}^M\sum\nolimits_{m'=1}^M\sqrt{\eta_{mk'}\eta_{m'k'}}{\rho}_{k}^2[n-\lambda]p_p{\rho}_{k}[\lambda-t_k^\text{d}]{\rho}_{k}[\lambda-t_k^\text{c}]\Biggl(\begin{array}{ll}\text{tr}(\boldsymbol{\Delta}_{mk}^\text{d}{\textbf{Z}}_{mk',d})\text{tr}(\boldsymbol{\Delta}_{m'k}^\text{c}{\textbf{Z}}_{m'k',c}^H)\\+\text{tr}(\boldsymbol{\Delta}_{mk}^\text{c}{\textbf{Z}}_{mk',c})\text{tr}(\boldsymbol{\Delta}_{m'k}^\text{d}{\textbf{Z}}_{m'k',d}^H)\end{array}\Biggr)\vspace{2 pt}
\\&\displaystyle+\sum\nolimits_{k'\in\mathcal{P}_k^\text{d}}\sum\nolimits_{m=1}^M\sum\nolimits_{m'=1}^M\sqrt{\eta_{mk'}\eta_{m'k'}}{\rho}_{k}^2[n-\lambda]p_p{\rho}_{k}^2[\lambda-t_k^\text{d}]\text{tr}(\boldsymbol{\Delta}_{mk}^\text{d}{\textbf{Z}}_{mk',d})\text{tr}(\boldsymbol{\Delta}_{m'k}^\text{d}{\textbf{Z}}_{m'k',d}^H)\vspace{2 pt}\\&\displaystyle+\sum\nolimits_{k'\in\mathcal{P}_k^\text{c}}\sum\nolimits_{m=1}^M\sum\nolimits_{m'=1}^M\sqrt{\eta_{mk'}\eta_{m'k'}}{\rho}_{k}^2[n-\lambda]p_p{\rho}_{k}^2[\lambda-t_k^\text{c}]\text{tr}(\boldsymbol{\Delta}_{mk}^\text{c}{\textbf{Z}}_{mk',c})\text{tr}(\boldsymbol{\Delta}_{m'k}^\text{c}{\textbf{Z}}_{m'k',c}^H).
\end{array}
 \vspace{-2 pt}
\label{DL_downlink}
   \end{equation}
   \hrulefill
   \end{figure*}
Meanwhile, the EMI power is given by
\begin{equation}
\begin{array}{ll}
\displaystyle\mathbb{E}\Big{\{}\Big{|}\sum\nolimits_{j=1}^{J}(\boldsymbol{\Phi}_j[n]\textbf{g}_{kj}^{\text{c}}[n])^T\textbf{n}_j[n]\Big{|}^2\Big{\}}\\\displaystyle=\sum\nolimits_{j=1}^{J}\mathbb{E}\Big{\{}\Big{|}(\boldsymbol{\Phi}_j\textbf{g}_{kj}^{\text{c}}[n])^T\textbf{n}_j[n]\Big{|}^2\Big{\}}\displaystyle=\sum\nolimits_{j=1}^{J}\beta_{kj}\sigma_j^2\text{tr}(\textbf{T}_j).
 \end{array}
 \label{EMI_downlink}
   \end{equation}  
\\
\textit{Proof:} Please refer to Appendix A.

\subsection{Sum SE Maximization}
In this section, we provide the RIS coefficient matrix design to
maximize the sum downlink SE. 
Note that the sum downlink SE depends on large-scale statistics, which experience slow variation since the distance variation is negligible compared with the initial distance. Thus, the proposed optimization problem can be performed across several coherence intervals\cite{9875036}. To this end, the optimization problem of the RIS coefficient matrix can be formulated as
 \begin{subequations}
    \begin{align}
      P_1:~&\mathop {\max }\limits_{{\boldsymbol{\Theta}}}~\text{SE}_{\text{sum}}=\sum\nolimits_{k=1}^K\text{SE}_k
      \\
      & \text{subject~to} \nonumber \\
      &|{\boldsymbol{\Phi}}_j(l,l)|=1,~\forall l,~\forall j,
    \end{align}
    \label{GD_optimization2}
  \end{subequations}
where $\boldsymbol{\Theta}=[\boldsymbol{\Phi}_1,...,\boldsymbol{\Phi}_J]\in\mathbb{C}^{JL\times L}$. This optimization problem is non-convex \cite{10297571,9875036,qian2024performanceanalysisstarrisassistedcellfree}, and to solve it in a suboptimal manner, we deploy the projected GA algorithm to obtain the local optimal solution of the maximization problem\cite{9875036,10373089}. 
For the proposed projected GA algorithm, increasing the objective from the current iteration ${\boldsymbol{\Phi}}_j^t,~\forall j$ to the gradient direction, has the following iterations\cite{10297571,10093070,10164189}
 \begin{equation}
\begin{array}{ll}
{\boldsymbol{\Phi}}_j^{t+1}=\text{P}_{{\boldsymbol{\Phi}}_j}\left({\boldsymbol{\Phi}}_j^t+\mu_j\nabla _{{\boldsymbol{\Phi}}_j^t}\text{SE}_{\text{sum}}({\boldsymbol{\Theta}}^t)\right),
     \end{array}
     \label{Optimization_1}
   \end{equation}
where $\mu_j$ is the step size for ${\boldsymbol{\Phi}}_j,~\forall j,$ and the superscript $t$ is the iteration index. To meet the constraints, we apply the projection functions $\text{P}_{{\boldsymbol{\Phi}}_j}\left({\boldsymbol{\Phi}_j}\right)$ following \cite{10297571}
\begin{equation}
    \begin{array}{cc}
[\text{P}_{{\boldsymbol{\Phi}}_j}\left({\boldsymbol{\Phi}}_j\right)]_{l,l}=\displaystyle\frac{{\boldsymbol{\Phi}}_j(l,l)}{|{\boldsymbol{\Phi}}_j(l,l)|}, ~l=1,...,L,~\forall j.
\end{array}
\end{equation}
  
Note that, referring to \cite{10418910}, the complex gradients of $\text{SE}_{\text{sum}}({\boldsymbol{\Theta}})$ in terms of ${\boldsymbol{\Phi}}_j$ can be computed as $\nabla _{{\boldsymbol{\Phi}}_j}\text{SE}_{\text{sum}}({\boldsymbol{\Theta}})$ with the decomposition as
  \begin{equation}
\begin{array}{ll}
\displaystyle \nabla _{{\boldsymbol{\Phi}}_j}\text{SE}_{\text{sum}}({\boldsymbol{\Theta}})  & \vspace{2 pt}\displaystyle=\nabla _{{\boldsymbol{\Phi}}_j}\displaystyle \sum\limits_{k=1}^K\sum\limits_{n=\lambda}^{\tau_c}\text{log}_2\left(1+\frac{S_k[n]}{I_k[n]}\right)\\ &=\displaystyle\sum\limits_{k=1}^K\sum\limits_{n=\lambda}^{\tau_c}\frac{\left(\nabla _{{\boldsymbol{\Phi}}_j}S_k[n]\right)I_k[n]-S_k[n]\nabla _{{\boldsymbol{\Phi}}_j}I_k[n]}{\text{ln}2I_k^2[n](1+\frac{S_k[n]}{I_k[n]})}.
     \end{array}
     \label{Optimization_4}
   \end{equation} 
According to \cite{Minka2000OldAN,4203075}, we can introduce the closed-form expressions of $\nabla _{{\boldsymbol{\Phi}}_j}S_k[n]$ as \eqref{Sk_derivate} at the top of this page with
\begin{figure*}[t!]
\vspace{-7 pt}
  \begin{equation}
\begin{array}{ll}
\nabla _{{\boldsymbol{\Phi}}_j}S_k[n]&\displaystyle= p_d\rho_k^2[n-\lambda]\nabla _{{\boldsymbol{\Phi}}_j}\left(\Big{|}\sum\nolimits_{m=1}^{M}\sqrt{\eta_{mk}}\text{tr}\left(\textbf{Q}_{mk}^\text{d}+\textbf{Q}_{mk}^\text{c}\right)\Big{|}^2\right)
\\&\displaystyle=
p_d\rho_k^2[n-\lambda]\sum\nolimits_{m=1}^{M}\sum\nolimits_{m'=1}^{M}\sqrt{\eta_{mk}\eta_{m'k}}\Bigg{(}\text{tr}\left(\textbf{Q}_{m'k}^\text{d}+\textbf{Q}_{m'k}^\text{c}\right)\nabla _{{\boldsymbol{\Phi}}_j}\text{tr}\left(\textbf{Q}_{mk}^\text{c}\right)+\text{tr}\left(\textbf{Q}_{mk}^\text{d}+\textbf{Q}_{mk}^\text{c}\right)\nabla _{{\boldsymbol{\Phi}}_j}\text{tr}\left(\textbf{Q}_{m'k}^\text{c}\right)\Bigg{)}\\&\displaystyle-
p_d\rho_k^2[n-\lambda]\sum\nolimits_{m=1}^{M}\sum\nolimits_{m'=1}^{M}{\eta_{mk}\eta_{m'k}}\text{tr}\left(\textbf{Q}_{m'k}^\text{d}+\textbf{Q}_{m'k}^\text{c}\right)\text{tr}\left(\textbf{Q}_{mk}^\text{d}+\textbf{Q}_{mk}^\text{c}\right)\nabla _{{\boldsymbol{\Phi}}_j}\left(\frac{1}{\sqrt{{\eta_{mk}\eta_{m'k}}}}\right),
  \end{array}
   \vspace{-3 pt}
     \label{Sk_derivate}
   \end{equation}
   \hrulefill
\end{figure*}
  \begin{equation}
\begin{array}{ll}
\displaystyle\nabla _{{\boldsymbol{\Phi}}_j}\left(\frac{1}{\sqrt{{\eta_{mk}\eta_{m'k}}}}\right)
=\frac{1}{2}\left(\begin{array}{ll}\displaystyle\sqrt{\frac{\eta_{mk}}{\eta_{m'k}}}\sum\nolimits_{k'=1}^K\nabla _{{\boldsymbol{\Phi}}_j}\text{tr}\left(\textbf{Q}_{mk'}^\text{c}\right)\\\displaystyle+\sqrt{\frac{\eta_{m'k}}{\eta_{mk}}}\sum\nolimits_{k'=1}^K\nabla _{{\boldsymbol{\Phi}}_j}\text{tr}\left(\textbf{Q}_{m'k'}^\text{c}\right)     \end{array}\right).
     \end{array}
     \label{Optimization_11}
   \end{equation}
\begin{figure*}[t!]
  \begin{equation}
\begin{array}{ll}
\nabla _{{\boldsymbol{\Phi}}_j}I_k[n]
&\displaystyle={p_d}\nabla _{{\boldsymbol{\Phi}}_j}\left( \sum\limits_{k'=1}^K\mathbb{E}\bigg{\{}\Big{|}\sum\limits_{m=1}^{M}\sqrt{\eta_{mk'}}\textbf{g}_{mk}^{T}[n]\hat{\textbf{g}}_{mk'}^*[\lambda]\Big{|}^2\bigg{\}}\right)-\nabla _{{\boldsymbol{\Phi}}_j}S_k[n]+\nabla _{{\boldsymbol{\Phi}}_j}\sum\limits_{j=1}^{J}\beta_{kj}\sigma_j^2\text{tr}(\textbf{T}_j)\\&\displaystyle=
p_d\sum\limits_{k'=1}^K\sum\limits_{m=1}^M\eta_{mk'}\left(
p_p\rho_{k'}^2[\lambda-t_{k'}^\text{c}]\left[\begin{array}{ll}\text{tr}\left((\mathbf{\Psi}_{mk'}^\text{c})^{-1}\mathbf{\Delta}_{mk'}^\text{c}\mathbf{\Delta}_{mk}\mathbf{\Pi}_{mk'j}\right)\\+\text{tr}\left(\mathbf{\Delta}_{mk}\mathbf{\Delta}_{mk'}^\text{c}(\mathbf{\Psi}_{mk'}^\text{c})^{-1}\mathbf{\Pi}_{mk'j}\right)\\-\sum\limits_{k''\in\mathcal{P}_{k'}^\text{c}}p_{k''}^\text{c}\text{tr}\left((\mathbf{\Psi}_{mk'}^\text{c})^{-1}\mathbf{\Delta}_{mk'}^\text{c}\mathbf{\Delta}_{mk}\mathbf{\Delta}_{mk'}^\text{c}(\mathbf{\Psi}_{mk'}^\text{c})^{-1}\mathbf{\Pi}_{mk'j}\right)\\-\beta_{mj}\sigma_j^2\text{tr}\left((\mathbf{\Psi}_{mk'}^\text{c})^{-1}\mathbf{\Delta}_{mk'}^\text{c}\mathbf{\Delta}_{mk}\mathbf{\Delta}_{mk'}^\text{c}(\mathbf{\Psi}_{mk'}^\text{c})^{-1}\textbf{R}_{mj,r}\right)
\end{array}
\right]+\text{tr}\left((\textbf{Q}_{mk'}^\text{d}+\textbf{Q}_{mk'}^\text{c})\mathbf{\Pi}_{mkj}\right)\right)\nabla _{{\boldsymbol{\Phi}}_j}\text{tr}(\textbf{T}_j)\\&\displaystyle-
p_d\sum\limits_{k'=1}^K\sum\limits_{m=1}^{M}\eta_{mk'}^2\text{tr}\left((\textbf{Q}_{mk'}^\text{d}+\textbf{Q}_{mk'}^\text{c})\mathbf{\Delta}_{mk}\right)\sum\limits_{k''=1}^K\nabla _{{\boldsymbol{\Phi}}_j}\text{tr}(\textbf{Q}_{mk''}^\text{c})

\\&\displaystyle+p_d\sum\limits_{k'\in\mathcal{P}_k^\text{c}}\sum\limits_{m=1}^{M}{\eta_{mk'}}\rho_k^2[n-\lambda]p_p\rho_k^2[\lambda-t_k^\text{c}]\left(
\begin{array}{ll}
   \beta_{mj}^2\beta_{kj}^2\text{tr}\left(\textbf{R}_{mj,r}\textbf{Z}_{mk',c}\textbf{R}_{mj,r}\textbf{Z}^H_{mk',c}\right)\cdot 2A_j^2\text{diag}\Big{(}{\textbf{R}}_j^T{\boldsymbol{\Phi}}_{j}^*{\textbf{R}}_j^T\textbf{T}_j^T\Big{)}\\+
   \sum\limits_{j'=1}^J\beta_{mj'}^2\beta_{kj'}^2\text{tr}(\textbf{T}_{j'}^2)\sqrt{p_p}\rho_{k'}[\lambda-t_{k'}]\mathbf{\Gamma}_{mmk'j'}\nabla _{{\boldsymbol{\Phi}}_j}\text{tr}(\textbf{T}_j)
\end{array}
\right)\\&\displaystyle-
p_d\sum\limits_{k'\in\mathcal{P}_k^\text{c}}\sum\limits_{m=1}^{M}{\eta_{mk'}^2}\rho_k^2[n-\lambda]p_p\rho_k^2[\lambda-t_k^\text{c}]
\sum\limits_{j=1}^J\beta_{mj}^2\beta_{kj}^2\text{tr}(\textbf{T}_j^2)\text{tr}(\textbf{R}_{mj,r}\textbf{Z}_{mk',c}\textbf{R}_{mj,r}\textbf{Z}^H_{mk',c})\sum\limits_{k''=1}^K\nabla _{{\boldsymbol{\Phi}}_j}\text{tr}(\textbf{Q}_{mk''}^\text{c})
\\&\displaystyle+p_d\sum\limits_{k'=1}^K\sum\limits_{m=1}^{M}\sum\limits_{m'=1}^{M}\sqrt{\eta_{mk'}\eta_{m'k'}}\left(
\begin{array}{ll}
\left(\sum\limits_{k''\in\mathcal{P}_{k'}^\text{c}}p_p\beta_{k''j}+\sigma^2_j\right)\beta_{mj}\beta_{m'j}\beta_{kj}\text{tr}\left(\textbf{R}_{mj,r}\textbf{Z}_{mk',c}\textbf{R}_{m'j,r}\textbf{Z}^H_{m'k',c}\right)\cdot 2A_j^2\text{diag}\Big{(}{\textbf{R}}_j^T{\boldsymbol{\Phi}}_{j}^*{\textbf{R}}_j^T\textbf{T}_j^T\Big{)}\\+
\sum\limits_{j'=1}^J\left(\sum\limits_{k''\in\mathcal{P}_{k'}^\text{c}}p_p\beta_{k''j'}+\sigma^2_{j'}\right)\beta_{mj'}\beta_{m'j'}\beta_{kj'}\text{tr}(\textbf{T}_{j'}^2)\sqrt{p_p}\rho_{k'}[\lambda-t_{k'}]\mathbf{\Gamma}_{mm'k'j'}\nabla _{{\boldsymbol{\Phi}}_j}\text{tr}(\textbf{T}_j)
\end{array}
\right)\\&\displaystyle-
p_d\sum\limits_{k'=1}^K\sum\limits_{m=1}^{M}\sum\limits_{m'=1}^{M}{\eta_{mk'}\eta_{m'k'}}\sum\limits_{j=1}^J\left(\sum\limits_{k''\in\mathcal{P}_{k'}^\text{c}}p_p\beta_{k''j}+\sigma^2_j\right)\beta_{mj}\beta_{m'j}\beta_{kj}\text{tr}(\textbf{T}_j^2)\text{tr}(\textbf{R}_{mj,r}\textbf{Z}_{mk',c}\textbf{R}_{m'j,r}\textbf{Z}^H_{m'k',c})\nabla _{{\boldsymbol{\Phi}}_j}\left(\frac{1}{\sqrt{{\eta_{mk'}\eta_{m'k'}}}}\right)
\\&\displaystyle+p_d\sum\limits_{k'\in\mathcal{P}_k^\text{d}\cap\mathcal{P}_k^\text{c}}\sum\limits_{m=1}^M\sum\limits_{m'=1}^M\sqrt{\eta_{mk'}\eta_{m'k'}}{\rho}_{k}^2[n-\lambda]p_p{\rho}_{k}[\lambda-t_k^\text{d}]{\rho}_{k}[\lambda-t_k^\text{c}]\Biggl(\begin{array}{ll}
\text{tr}(\boldsymbol{\Delta}_{mk}^\text{d}{\textbf{Z}}_{mk',d})\nabla _{{\boldsymbol{\Phi}}_j}\left(\text{tr}(\boldsymbol{\Delta}_{m'k}^\text{c}{\textbf{Z}}_{m'k',c}^H)\right)\\+\text{tr}(\boldsymbol{\Delta}_{m'k}^\text{d}{\textbf{Z}}_{m'k',d}^H)\nabla _{{\boldsymbol{\Phi}}_j}\left(\text{tr}(\boldsymbol{\Delta}_{mk}^\text{c}{\textbf{Z}}_{mk',c})\right)\end{array}\Biggr)\\&\displaystyle-
p_d\sum\limits_{k'\in\mathcal{P}_k^\text{d}\cap\mathcal{P}_k^\text{c}}\sum\limits_{m=1}^M\sum\limits_{m'=1}^M{\eta_{mk'}\eta_{m'k'}}{\rho}_{k}^2[n-\lambda]p_p{\rho}_{k}[\lambda-t_k^\text{d}]{\rho}_{k}[\lambda-t_k^\text{c}]\Biggl(\begin{array}{ll}\text{tr}(\boldsymbol{\Delta}_{mk}^\text{d}{\textbf{Z}}_{mk',d})\text{tr}(\boldsymbol{\Delta}_{m'k}^\text{c}{\textbf{Z}}_{m'k',c}^H)\\+\text{tr}(\boldsymbol{\Delta}_{mk}^\text{c}{\textbf{Z}}_{mk',c})\text{tr}(\boldsymbol{\Delta}_{m'k}^\text{d}{\textbf{Z}}_{m'k',d}^H)\end{array}\Biggr)\nabla _{{\boldsymbol{\Phi}}_j}\left(\frac{1}{\sqrt{{\eta_{mk'}\eta_{m'k'}}}}\right)

\\&\displaystyle+p_d\sum\limits_{k'\in\mathcal{P}_k^\text{c}}\sum\limits_{m=1}^M\sum\limits_{m'=1}^M\sqrt{\eta_{mk'}\eta_{m'k'}}{\rho}_{k}^2[n-\lambda]p_p{\rho}_{k}^2[\lambda-t_k^\text{c}]\Biggl(\text{tr}(\boldsymbol{\Delta}_{mk}^\text{c}{\textbf{Z}}_{mk',c})\nabla _{{\boldsymbol{\Phi}}_j}\left(\text{tr}(\boldsymbol{\Delta}_{m'k}^\text{c}{\textbf{Z}}_{m'k',c}^H)\right)+\text{tr}(\boldsymbol{\Delta}_{m'k}^\text{c}{\textbf{Z}}_{m'k',c}^H)\nabla _{{\boldsymbol{\Phi}}_j}\left(\text{tr}(\boldsymbol{\Delta}_{mk}^\text{c}{\textbf{Z}}_{mk',c})\right)\Biggr)\\&\displaystyle-

p_d\sum\limits_{k'\in\mathcal{P}_k^\text{c}}\sum\limits_{m=1}^M\sum\limits_{m'=1}^M{\eta_{mk'}\eta_{m'k'}}{\rho}_{k}^2[n-\lambda]p_p{\rho}_{k}^2[\lambda-t_k^\text{c}]\text{tr}(\boldsymbol{\Delta}_{mk}^\text{c}{\textbf{Z}}_{mk',c})\text{tr}(\boldsymbol{\Delta}_{m'k}^\text{c}{\textbf{Z}}_{m'k',c}^H)\nabla _{{\boldsymbol{\Phi}}_j}\left(\frac{1}{\sqrt{{\eta_{mk'}\eta_{m'k'}}}}\right)

\\&\displaystyle-\nabla _{{\boldsymbol{\Phi}}_j}S_k[n]+\beta_{kj}\sigma_j^2\nabla _{{\boldsymbol{\Phi}}_j}\text{tr}(\textbf{T}_j),
  \end{array} \vspace{-5 pt}
     \label{Ik_derivate}
   \end{equation} 
         \hrulefill
\end{figure*}
\begin{figure*}[t!]
\vspace{-9 pt}
  \begin{equation}
\begin{array}{ll}
\displaystyle\mathbf{\Gamma}_{mm'k'j'}\displaystyle=\text{tr}\left((\mathbf{\Psi}_{mk'}^\text{c})^{-1}\textbf{R}_{m'j',r}\textbf{Z}_{m'k',c}^H\textbf{R}_{mj',r}\mathbf{\Pi}_{mk'j}\right)+\text{tr}\left(\textbf{R}_{mj',r}\textbf{Z}_{mk',c}\textbf{R}_{m'j',r}(\mathbf{\Psi}_{m'k'}^\text{c})^{-1}\mathbf{\Pi}_{m'k'j}\right)\\~~~~~~~~~\displaystyle-\sum\limits_{k''\in\mathcal{P}_{k'}^\text{c}}p_p\text{tr}\left((\mathbf{\Psi}_{mk'}^\text{c})^{-1}\textbf{R}_{m'j',r}\textbf{Z}_{m'k',c}^H\textbf{R}_{mj',r}\mathbf{\Delta}_{mk'}^\text{c}(\mathbf{\Psi}_{mk'}^\text{c})^{-1}\mathbf{\Pi}_{mk''j}\right)-\beta_{mj}\sigma_j^2\text{tr}\left((\mathbf{\Psi}_{mk'}^\text{c})^{-1}\textbf{R}_{m'j',r}\textbf{Z}_{m'k',c}^H\textbf{R}_{mj',r}\mathbf{\Delta}_{mk'}^\text{c}(\mathbf{\Psi}_{mk'}^\text{c})^{-1}\textbf{R}_{mj,r}\right)\\~~~~~~~~~\displaystyle-\sum\limits_{k''\in\mathcal{P}_{k'}^\text{c}}p_p\text{tr}\left((\mathbf{\Psi}_{m'k'}^\text{c})^{-1}\mathbf{\Delta}_{m'k'}^\text{c}\textbf{R}_{mj',r}\textbf{Z}_{mk',c}\textbf{R}_{m'j',r}(\mathbf{\Psi}_{m'k'}^\text{c})^{-1}\mathbf{\Pi}_{m'k''j}\right)-\beta_{mj}\sigma_j^2\text{tr}\left((\mathbf{\Psi}_{m'k'}^\text{c})^{-1}\mathbf{\Delta}_{m'k'}^\text{c}\textbf{R}_{mj',r}\textbf{Z}_{mk',c}\textbf{R}_{m'j',r}(\mathbf{\Psi}_{m'k'}^\text{c})^{-1}\textbf{R}_{m'j,r}\right),\end{array} \vspace{-5 pt}
   \end{equation} 
         \hrulefill
\end{figure*}
\begin{figure*}[t!]
\vspace{-9 pt}
  \begin{equation}
\begin{array}{ll}
\nabla _{{\boldsymbol{\Phi}}_j}\left(\text{tr}(\boldsymbol{\Delta}_{mk}^\text{c}{\textbf{Z}}_{mk',c}^H)\right)=\text{tr}({\textbf{Z}}_{mk',c}^H\mathbf{\Pi}_{mkj})\nabla _{{\boldsymbol{\Phi}}_j}\text{tr}(\textbf{T}_j)+\sqrt{p_p}\rho_{k'}[\lambda-t_{k'}^\text{c}]\left[\begin{array}{ll}\text{tr}\left(\mathbf{\Delta}_{mk}^\text{c}(\mathbf{\Psi}_{mk'}^\text{c})^{-1}\mathbf{\Pi}_{mk'j}\right)\\-\sum\nolimits_{k''\in\mathcal{P}_{k'}^\text{c}}p_p\text{tr}\left((\mathbf{\Psi}_{mk'}^\text{c})^{-1}\mathbf{\Delta}_{mk'}^\text{c}\mathbf{\Delta}_{mk}^\text{c}(\mathbf{\Psi}_{mk'}^\text{c})^{-1}\mathbf{\Pi}_{mk''j}\right)\\-\beta_{mj}\sigma_j^2\text{tr}\left((\mathbf{\Psi}_{mk'}^\text{c})^{-1}\mathbf{\Delta}_{mk'}^\text{c}\mathbf{\Delta}_{mk}^\text{c}(\mathbf{\Psi}_{mk'}^\text{c})^{-1}\textbf{R}_{mj,r}\right)
\end{array}
\right]\nabla _{{\boldsymbol{\Phi}}_j}\text{tr}(\textbf{T}_j)
,
\end{array} \vspace{-5 pt}
   \end{equation} 
         \hrulefill
\end{figure*}
\begin{figure*}[t!]
%\vspace{-6 pt}
  \begin{equation}
\begin{array}{ll}
\nabla _{{\boldsymbol{\Phi}}_j}\left(\text{tr}(\boldsymbol{\Delta}_{mk}^\text{c}{\textbf{Z}}_{mk',c})\right)=\text{tr}({\textbf{Z}}_{mk',c}\mathbf{\Pi}_{mkj})\nabla _{{\boldsymbol{\Phi}}_j}\text{tr}(\textbf{T}_j)+\sqrt{p_p}\rho_{k'}[\lambda-t_{k'}^\text{c}]\left[\begin{array}{ll}\text{tr}\left((\mathbf{\Psi}_{mk'}^\text{c})^{-1}\mathbf{\Delta}_{mk}^\text{c}\mathbf{\Pi}_{mk'j}\right)\\-\sum\nolimits_{k''\in\mathcal{P}_{k'}^\text{c}}p_p\text{tr}\left((\mathbf{\Psi}_{mk'}^\text{c})^{-1}\mathbf{\Delta}_{mk}^\text{c}\mathbf{\Delta}_{mk'}^\text{c}(\mathbf{\Psi}_{mk'}^\text{c})^{-1}\mathbf{\Pi}_{mk''j}\right)\\-\beta_{mj}\sigma_j^2\text{tr}\left((\mathbf{\Psi}_{mk'}^\text{c})^{-1}\mathbf{\Delta}_{mk}^\text{c}\mathbf{\Delta}_{mk'}^\text{c}(\mathbf{\Psi}_{mk'}^\text{c})^{-1}\textbf{R}_{mj,r}\right)
\end{array}
\right]\nabla _{{\boldsymbol{\Phi}}_j}\text{tr}(\textbf{T}_j).
\end{array} \vspace{-3 pt}
     \label{Ik_derivate4}
   \end{equation} 
         \hrulefill
\end{figure*}
Meanwhile, $\nabla _{{\boldsymbol{\Phi}}_j}I_k[n]$ are given by \eqref{Ik_derivate}-\eqref{Ik_derivate4} at the top of the following pages. Based on the above-mentioned observations, Algorithm \ref{Algorithm} delivers the iterative procedure of the projected GA algorithm to converge to a stationary point of $P_1$\cite{10297571}. We initialize the step size $\mu_j,~\forall j,$ and reduce it by increasing the iterations. Note that we introduce the suitable step size at each
iteration based on $\nabla _{{\boldsymbol{\Phi}}_j}\text{SE}_{\text{sum}}({\boldsymbol{\Theta}})$ for each RIS to further improve the performance of the projected GA algorithm.
   \\
\textit{Proof:} Please refer to Appendix B.
  \begin{algorithm}[t!]
      \caption{Projected GA Algorithm Based RIS Design} 
\begin{algorithmic}[1]
\renewcommand{\algorithmicrequire}{\textbf{Inputs:}}
\Require

$\epsilon$ (tolerance), IterMax; 
\renewcommand{\algorithmicensure}{\textbf{Output:}}
\Ensure
${\boldsymbol{\Phi}}_j$, $~\forall j$, $\text{SE}_{\text{sum}}({\boldsymbol{\Theta}})$
\State Initialize ${\boldsymbol{\Phi}}^0_j$, $\textbf{f}^0=\text{SE}_{\text{sum}}({\boldsymbol{\Theta}}^0),~\forall j$
\For{$\text{i}= 1 :\text{IterMax}$}
\State ${\boldsymbol{\Phi}}_j=[],~\forall j$
\State $\mu_j=(1-\displaystyle\frac{\text{i}-1}{2\cdot\text{IterMax}})/{|\nabla _{{\boldsymbol{\Phi}}_j^0}\text{SE}({\boldsymbol{\Theta}}^0)|}$;~~~~(step size)
\State  ${\boldsymbol{\Phi}}_j=\text{P}_{{\boldsymbol{\Phi}}_j}\left({\boldsymbol{\Phi}}_j^0+\mu_j\nabla _{{\boldsymbol{\Phi}}_j^0}\text{SE}_{\text{sum}}({\boldsymbol{\Theta}}^0)\right),~\forall j$
\State Calculate $\textbf{f}=\text{SE}_{\text{sum}}({\boldsymbol{\Theta}}),~\forall j$
\If {$\big{|}\textbf{f}-\textbf{f}^0\big{|}\leq \epsilon$}
\State $\textbf{break}$
       \Else  
       \State
       ${\boldsymbol{\Phi}}_j^0={\boldsymbol{\Phi}}_j$, $\textbf{f}^0=\textbf{f},~\forall j$;
\EndIf
\EndFor
\end{algorithmic}
\label{Algorithm}
  \end{algorithm}
\section{Numerical Results}

%\vspace{-8 pt}
We shall consider a simulation setup with APs, users and RISs located within a geographic area of $D\times D$ km${^2}$, where $D=0.5$ km. Referring to \cite{10621117}, APs and users/RISs are randomly and independently distributed in two adjacent sub-regions with $x^{\text{AP}},y^{\text{AP}}\in\left[-\frac{D}{2},0\right]$ km, $x^{\text{user}},y^{\text{user}}\in\left[0,\frac{D}{2}\right]$ km and $x^{\text{RIS}},y^{\text{RIS}}\in\left[0,\frac{D}{2}\right]$ km. $\beta_x=\text{PL}_x\cdot z_x$ 
($x=mk,~mj,~kj$) models the large-scale fading coefficients. $\text{PL}_x$ is the three-slope path loss and $z_x$ is log-normal shadowing with the standard deviation $\sigma_{\text{sh}}$ \cite{7827017}. Likewise, $d_0=10$
m, $d_1=50$ m and $\sigma_{\text{sh}}=8~\text{dB} $. The AP, RIS and user heights are 15 m, 30 m and 1.65 m \cite{10225319}. Moreover, the exponential correlation model in\cite{951380,7500452} is utilized for the spatial correlation of APs. Unless otherwise stated, all users experience the same velocity, $v_k=v,~\forall k$. The pilot sequence length is $\tau_p=3$, the carrier frequency equals $f_c=1.9$ GHz,
the bandwidth equals $B=20$ MHz, the time instant length equals $T_s=0.01$ ms. We also set downlink power as $p_{d}=23~\text{dBm}$, pilot and uplink power as $p_{p}=p_u=20~\text{dBm}$ and $\sigma^2=-91$ dBm\cite{10264149}. We assume that all RIS elements provide $d_h=d_v=\lambda/4$\cite{10167480,9665300}. For the projected GA algorithm, we set the maximum number of iterations IterMax$=20$ and tolerance $\epsilon=10^{-6}$\cite{10164189}.

\subsection{Channel Estimation Accuracy}
\begin{figure}[t!]
		\centering
\includegraphics[width=0.82\columnwidth]{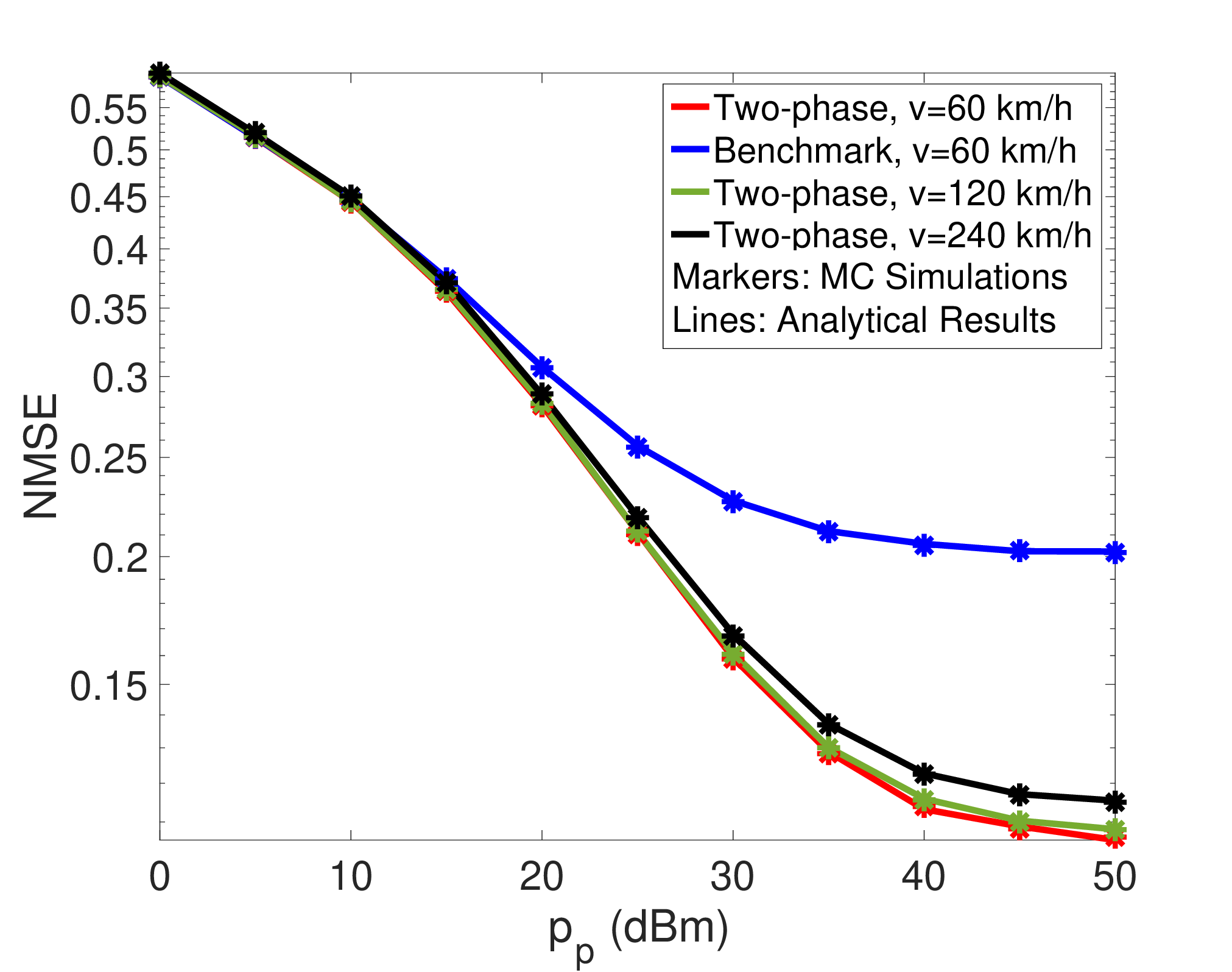} 
		\caption{NMSE vs $p_p$ with $\varsigma=20~\text{dB}$, $M=10$, $N=2$, $K=10$, $J=4$, $L=16$.}
		\label{NMSE}
        \vspace{-5 pt}
\end{figure}
\begin{figure}[t!]
		\centering
\includegraphics[width=0.82\columnwidth]{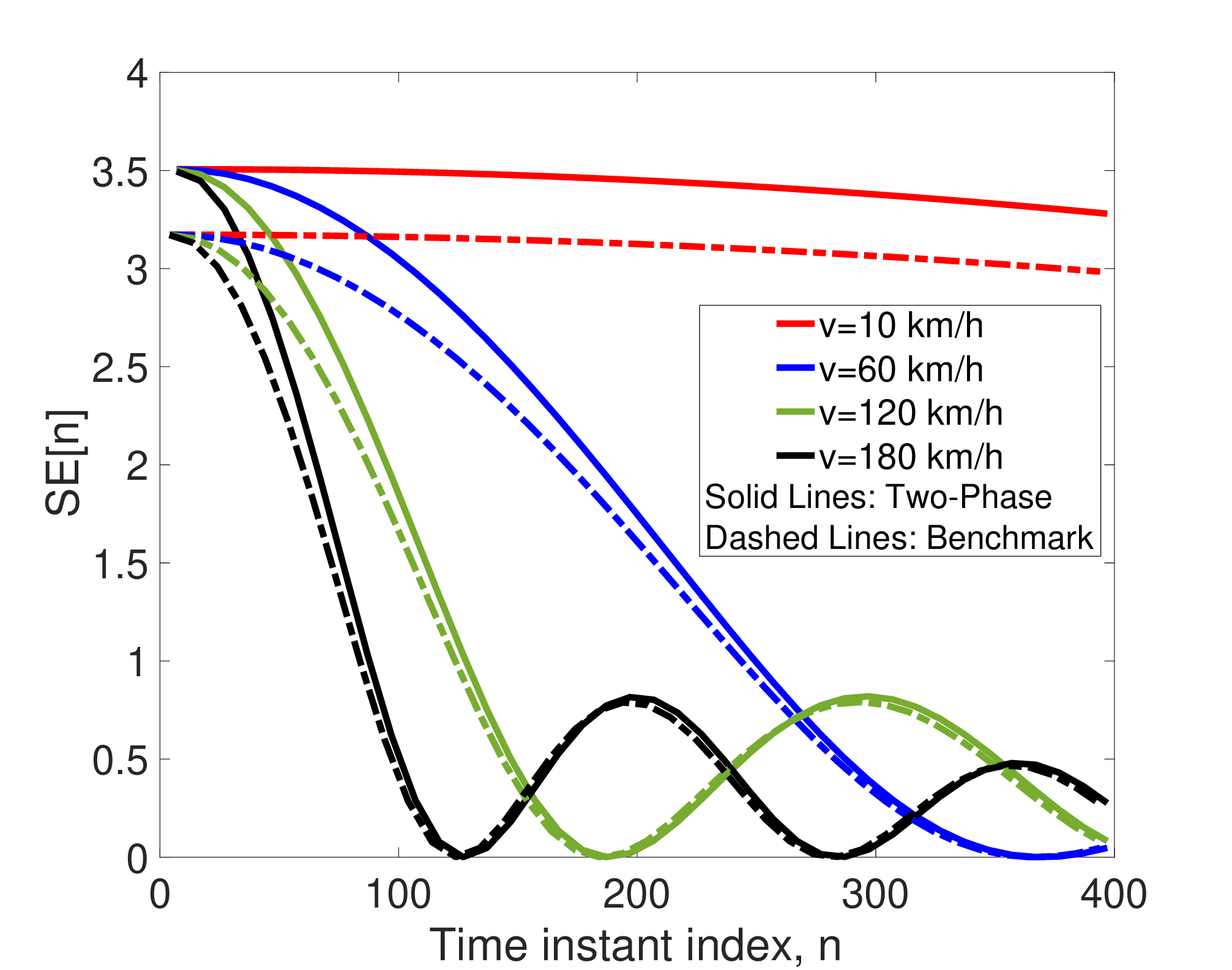} 
		\caption{SE vs the time instant index with $\varsigma=20~\text{dB}$, $M=10$, $N=2$, $K=10$, $J=4$, $L=16$.}
	\label{resouce_block_length}
    \vspace{-5 pt}
        \end{figure}
The NMSEs of the proposed system versus $p_p$ with different user velocities are shown in Fig. \ref{NMSE}. For comparison, the conventional MMSE scheme with randomly assigned pilots, which estimates the aggregate channels simultaneously, shall serve as the benchmark scheme \cite{10468556,10167480,9875036}. Meanwhile, the pilot length for the benchmark scheme is $\tau_p$. The analytical closed-form expressions generated by \eqref{D_est1}-\eqref{D_est2}, \eqref{R_est1}-\eqref{Psi_c} and \eqref{aggregate_uplink_channel_estimation}-\eqref{NMSE_expression} closely match the Monte Carlo (MC) simulations. 
It shows that our novel two-phase channel estimation scheme improves the accuracy of the estimation, giving a lower NMSE when pilot resources are limited. For example,  when $v=60 $ km/h and $p_p=20$ dBm, the two-phase scheme achieves a nearly $3\%$-likely improvement in NMSE than the benchmark scheme, and the improvement trend grows up to around $90\%$-likely improvement when $p_p=50$ dBm. In this case, the proposed two-phase scheme can bloom its benefits with larger pilot power. Note that $\tau_p=3 <K$ introduces pilot contamination, and EMI diminishes estimation accuracy. Thus, a non-zero error floor appears as $p_p$ increases. Moreover, increasing user velocity contributes to greater channel aging effects and reduces temporal correlation coefficients, further degrading estimation accuracy. These results indicate that the two-phase estimation scheme significantly improves channel estimation accuracy, outperforming the benchmark scheme.

\subsection{Channel Aging-aware Resource Block Length}
Fig. \ref{resouce_block_length} shows the downlink $\text{SE}[n]=\sum\nolimits_{k=1}^{K}\text{log}_2\Big{(} 1+\text{SINR}_k[n]\Big{)}$ at $n$-th time instant regarding the first 400 time instant index. The downlink data transmission for the proposed two-phase scheme starts at $n=2\tau_p+1$, while the benchmark scheme starts at $n=\tau_p+1$.
Although the two-phase scheme exhibits a reduced transmission efficiency compared to the benchmark scheme by $(\tau_c-2\tau_p)/(\tau_c-\tau_p)$, 
the results reveal that the proposed two-phase scheme outperforms the benchmark scheme at each time instant. We can also find that the two-phase scheme can achieve a nearly $10\%$-likely SE improvement compared to the benchmark scheme. As the temporal correlation coefficient diminishes with increasing time instant index, $\text{SE}[n]$ decreases. Meanwhile, the first zero position shifts leftward with higher user velocity, and the fluctuation peak becomes smaller with increasing time instant index. Therefore, selecting a reasonable resource block length is essential to mitigate channel aging effects. In the following, we select $\tau_c$ not to exceed the first zero index value at $v = 120 ~\text{km/h}$. As such, we consider $\tau_c=196$, and all users are assumed to have the same velocity $v\leq 120 ~\text{km/h}$.

\subsection{Spectral Efficiency Analysis}
\begin{figure}[t!]
		\centering	\includegraphics[width=0.82\columnwidth]{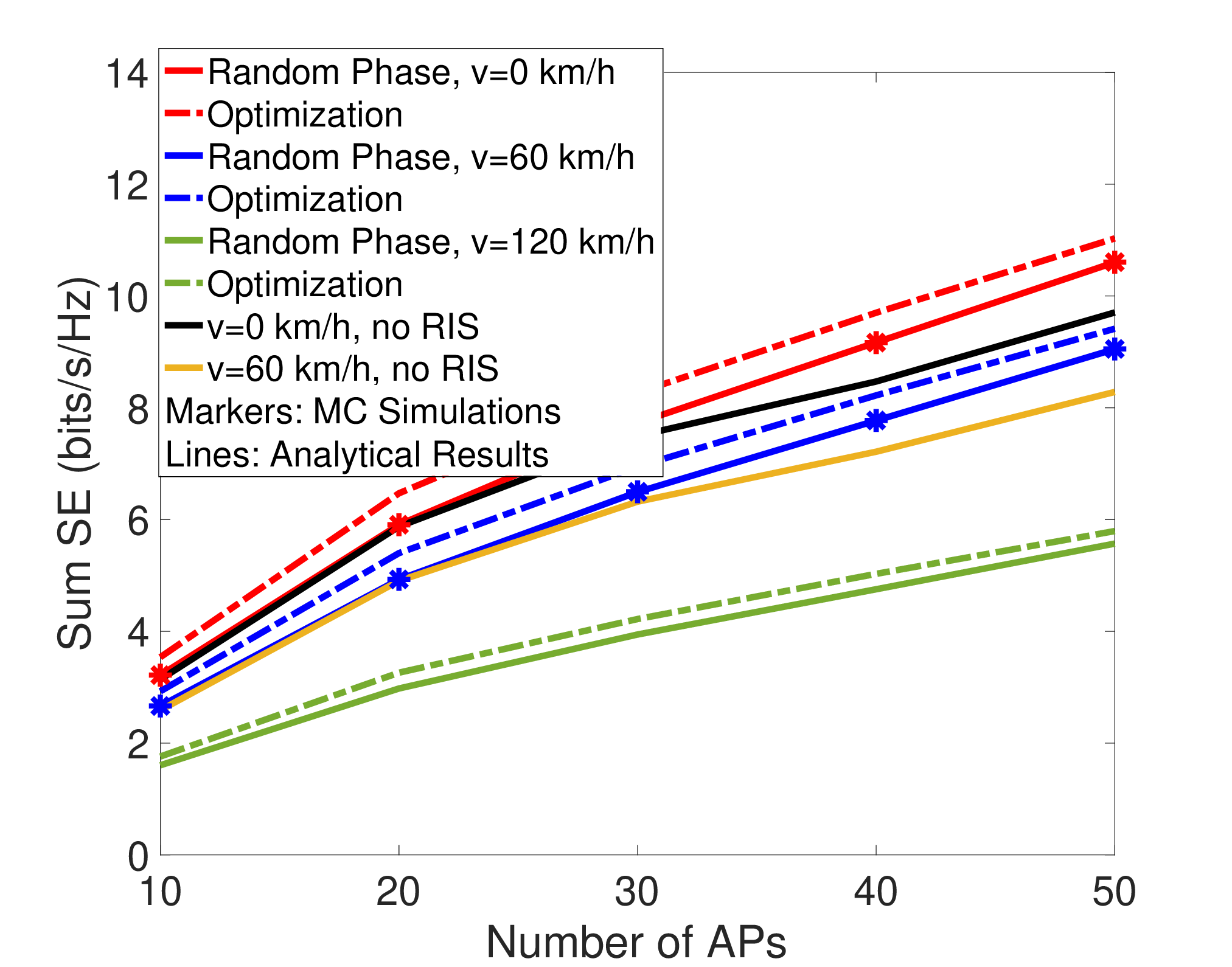} 
		\caption{Sum SE vs the number of APs with $\varsigma=20~\text{dB}$, $N=2$, $K=10$, $J=4$, $L=16$.}
		\label{VaryingM}
        \vspace{-5 pt}
        \end{figure}
\begin{figure}[t!]
		\centering
\includegraphics[width=0.82\columnwidth]{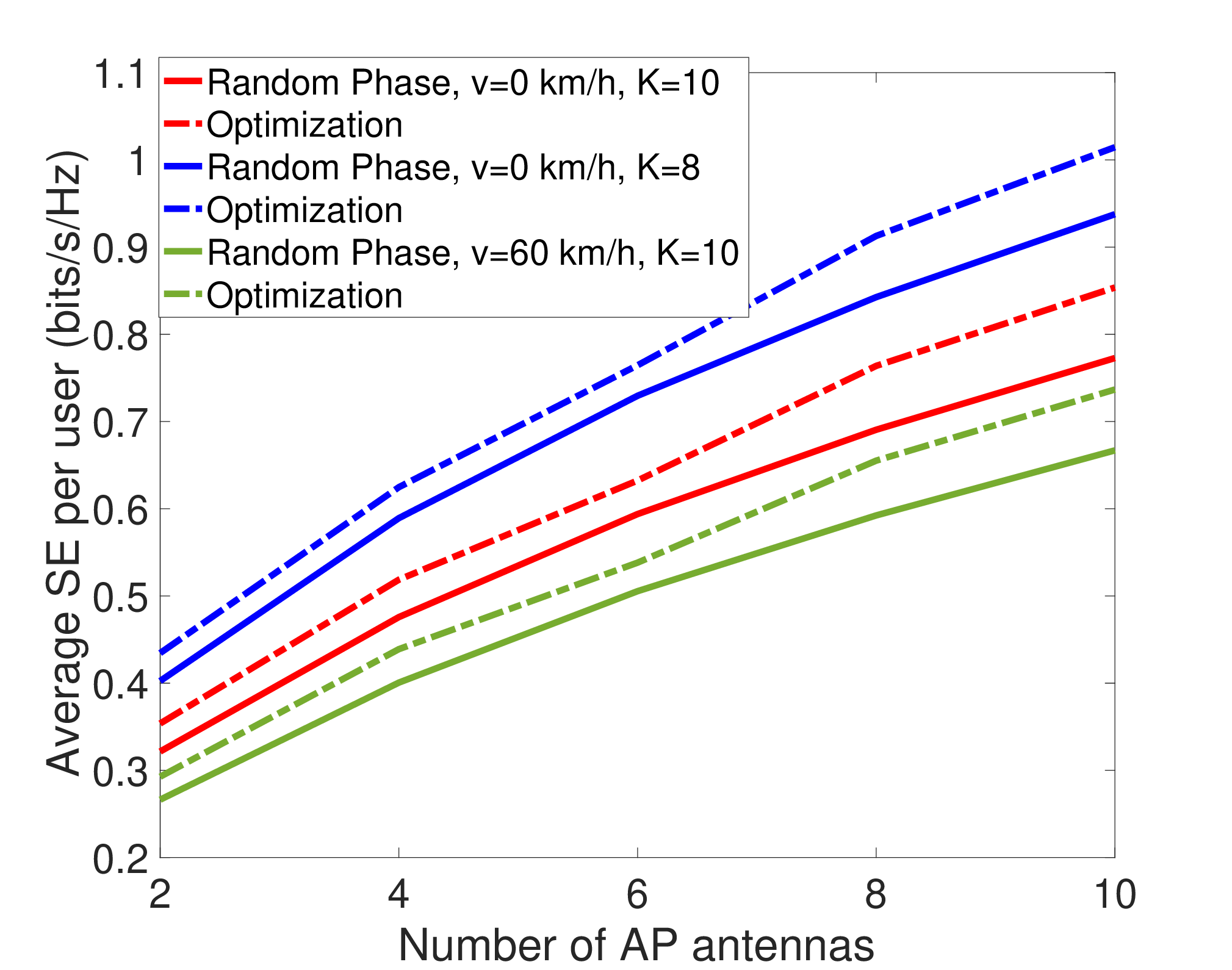} 
	\caption{Average SE vs the number of antennas per AP with $\varsigma=20~\text{dB}$, $M=10$, $J=4$, $L=16$.}
		\label{VaryingN}
        \vspace{-5 pt}
        \end{figure}
\begin{figure}[t!]
		\centering
\includegraphics[width=0.82\columnwidth]{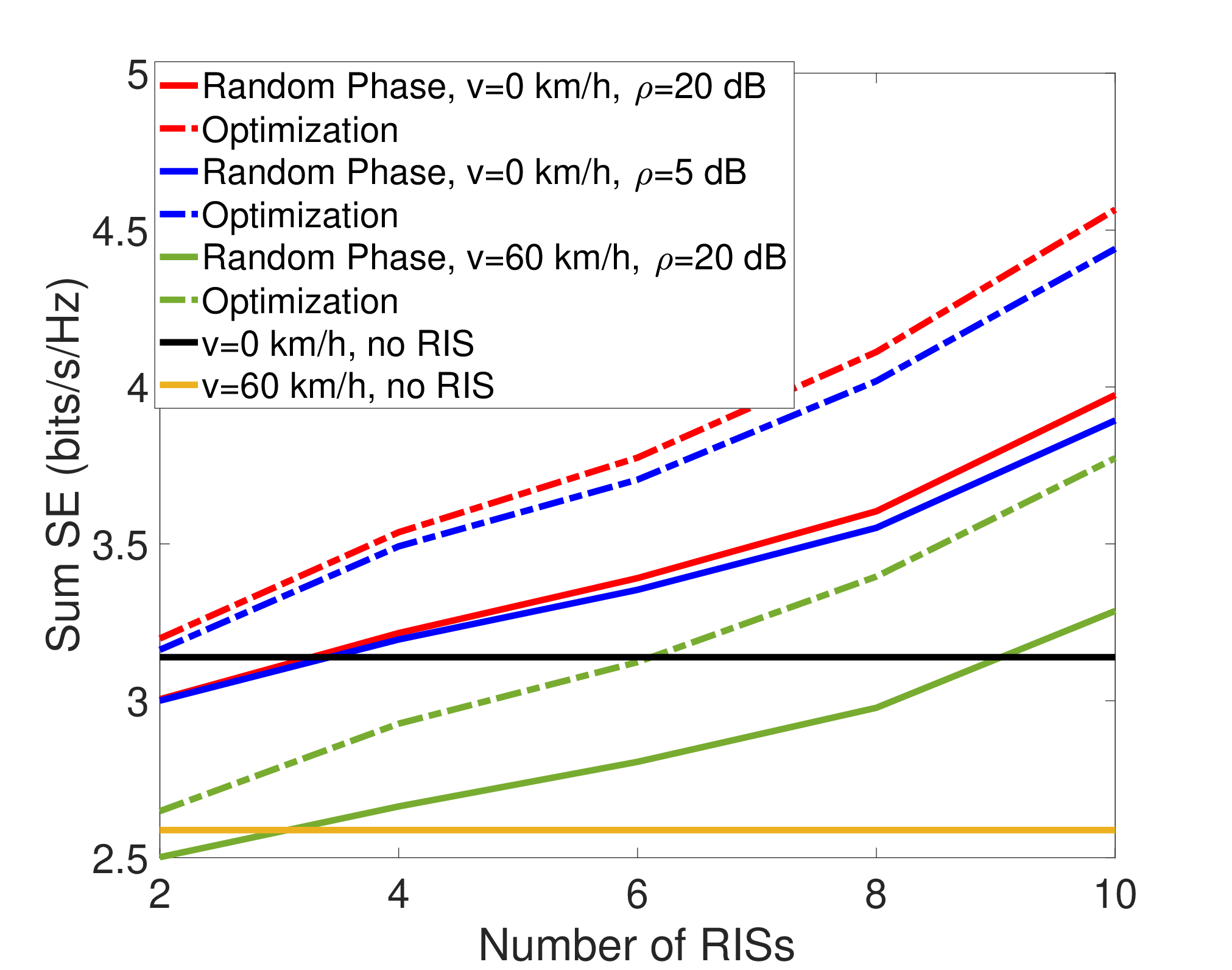} 
		\caption{Sum SE vs the number of RIS with $M=10$, $N=2$, $K=10$, $L=16$.}
		\label{VaryingJ}\vspace{-5 pt}
\end{figure}
\begin{figure}[t!]
		\centering
\includegraphics[width=0.82\columnwidth]{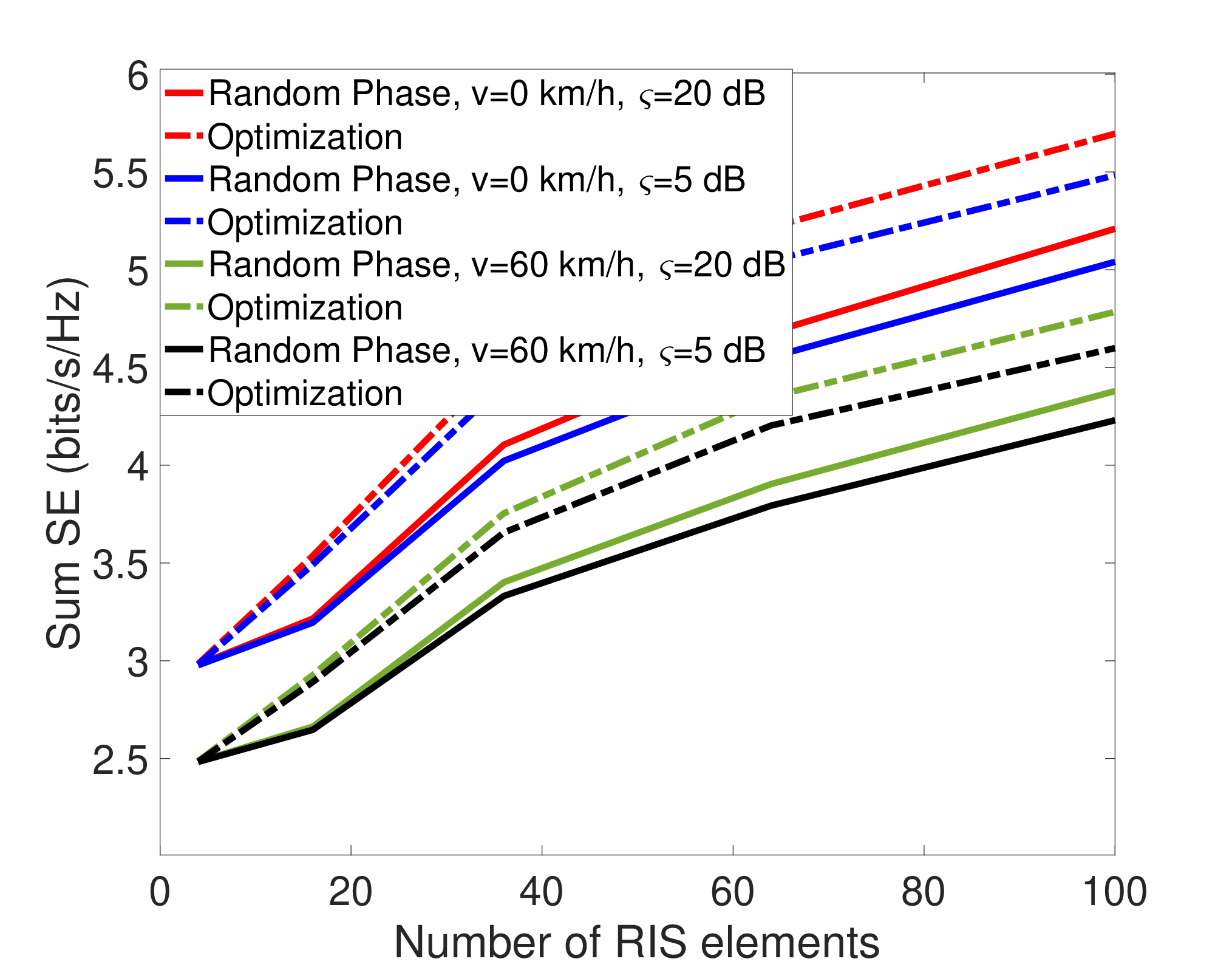} 
		\caption{Sum SE vs the number of elements per RIS with $M=10$, $N=2$, $K=10$, $J=4$.}
		\label{VaryingL}\vspace{-3 pt}
\end{figure}

Fig. \ref{VaryingM} depicts the sum SE, denoted as $~\text{SE}_{\text{sum}}=\sum\nolimits_{k=1}^K\text{SE}_k$, versus the number of APs and Fig. \ref{VaryingN} shows the average SE per user, denoted as $~\text{SE}_{\text{ave}}=\sum\nolimits_{k=1}^K\text{SE}_k/K$ as a function of the number of antennas per AP. Different user velocities and number of users are evaluated. For comparison, the RIS-free cell-free massive MIMO system utilizes the benchmark channel estimation scheme outlined in this work. In Fig. \ref{VaryingM}, the analytical closed-form expressions generated by \eqref{downlink_SE_description}-\eqref{EMI_downlink} can closely match the MC simulations. The results reveal that increasing the number of APs and antennas per AP can enhance spatial degrees of freedom to facilitate more efficient beamforming; therefore, the sum SE can be improved. In Fig. \ref{VaryingM}, the sum downlink SE optimized by the proposed projected GA algorithm exhibits over a $5\%\sim10\%$-likely SE improvement than that with the random RIS phases. Moreover, increasing APs can compensate for the performance degradation caused by channel aging and EMI since increasing APs can achieve an SINR increase nearly proportional to it.
Fig. \ref{VaryingN} indicates that the sum SE achieved by the proposed projected GA algorithm surpasses that with random RIS phases by approximately $10\%$ when $M=4$. However, increasing the number of users may diminish average performance since more users introduce intensified pilot contamination and inter-user interference. Note that increasing the number of AP antennas can alleviate this decrease. Although EMI and channel aging introduce SE degradation, the proposed RIS-assisted system outperforms RIS-free cell-free massive MIMO systems, introducing a nearly $10\%$-likely SE improvement. In this case, integrating RISs into cell-free massive MIMO systems is advantageous. Meanwhile, an appropriate increase in the number of APs and antennas per AP is required to enhance system performance.

Fig. \ref{VaryingJ} and Fig. \ref{VaryingL} illustrate the sum SE as a function of the number of RISs and the number of elements per RIS, respectively. Increasing the number of RISs and elements per RIS can effectively neutralize the performance degradation caused by channel aging and moderate EMI. For example, the proposed system achieves a notable SE improvement of $25\%$ when $\varsigma=20$ dB and $J=10$ compared to the benchmark RIS-free cell-free massive MIMO system. Passive beamforming and proper RIS configuration can improve interference management to release the necessity of introducing RISs and finding the RIS settings. Compared to the random RIS phases, the proposed projected GA algorithm can yield an extra $10\%\sim15\%$-likely SE improvement with an increasing number of RISs. Similarly, it can introduce a nearly $10\%$-likely SE improvement with an increasing number of RIS elements. These results underscore the effectiveness of the RIS coefficient matrix optimization
algorithm in improving RIS-assisted system performance. 
However, under severe EMI conditions, i.e., $\varsigma=5 $ dB, the proposed system struggles to deliver effective performance improvements, as the increased RISs and RIS elements exacerbate EMI to reduce system performance improvement. 
The result shows that EMI is non-negligible, and introducing EMI elimination schemes is necessary to further reap the benefits of deploying RISs. Despite the adverse effects of severe EMI, the advantages of incorporating RISs into cell-free massive MIMO systems remain substantial, and the RIS coefficient matrix design is necessary to compensate for the performance reduction caused by EMI.

\section{Conclusion}
This paper analyzed the performance of RIS-assisted cell-free massive MIMO systems experiencing channel aging and EMI. To our knowledge, this work is the first to establish the system model for spatially correlated RIS-assisted cell-free massive MIMO systems with the joint effect of channel aging and EMI. Furthermore, it provides analytical expressions for downlink SE performance analysis. We applied a novel two-phase estimation scheme to mitigate the effects of channel aging and EMI. We adopted the downlink conjugate beamforming with fractional power control for performance improvement, deriving closed-form expressions for downlink SE. Subsequently, we utilized the closed-form expressions to maximize the sum SE with respect to the RIS coefficient matrices via the projected GA algorithm. It was shown that our novel two-phase channel estimation scheme could reduce the performance degradation caused by channel aging and EMI. Meanwhile, the
proposed RIS coefficient matrix optimization could also alleviate performance degradation. However, the benefits of introducing more RISs and RIS elements could be diminished in environments with severe EMI, posing challenges for effective deployment. 

\ifCLASSOPTIONcaptionsoff
  \newpage
\fi

\begin{appendices}

\section
{Derivation of uplink SE approximations}
This appendix provides the detailed derivation of $\text{SINR}_k[n]$ by using the UatF bound \cite{8845768,9416909} to obtain the SE of $k$-th user in \eqref{downlink_SINR}. According to the MMSE properties, the estimate $\hat{\textbf{g}}_{mk}[\lambda]$ and estimation error $\tilde{\textbf{g}}_{mk}[\lambda]$ are uncorrelated\cite{9875036}. The users in the same set, $k'\in\mathcal{P}_k^\text{x}$, $\text{x}=\text{c},\text{d}$, share the same pilot sequence and $\hat{\textbf{g}}_{mk'}[\lambda]$ is correlated with ${\textbf{g}}_{mk}[\lambda]$. We can obtain the following derivations referring to \cite{10468556,10167480}.
\subsubsection{Compute $\mathbb{E}\{|S_k[n]|^2\}$} the desired signal is given by,
\begin{equation}
\setlength{\abovedisplayskip}{6pt}
\begin{array}{ll}
\displaystyle\mathbb{E}\{|S_k[n]|^2\}\displaystyle=\mathbb{E}\Big{\{}{p_d}\rho_k^2[n-\lambda]\Big{|}\sum\nolimits_{m=1}^{M}\sqrt{\eta_{mk}}\mathbb{E}\bigg{\{}\textbf{g}_{mk}^{T}[\lambda]\hat{\textbf{g}}_{mk}^*[\lambda]\bigg{\}}\Big{|}^2\Big{\}}
\vspace{2 pt}\\~~~~~~~~~~~~~\displaystyle={p_d}\rho_k^2[n-\lambda]\Big{|}\sum\nolimits_{m=1}^{M}\sqrt{\eta_{mk}}\text{tr}\left(\textbf{Q}_{mk}^\text{d}+\textbf{Q}_{mk}^\text{c}\right)\Big{|}^2.
\end{array}
   \end{equation}

\subsubsection{Compute $\mathbb{E}\{|I_k[n]|^2\}$} We can first decompose the inter-user interference term as
\begin{equation}
\begin{aligned}
&\displaystyle\mathbb{E}\Big{\{}|\text{UI}_{\text{kk'}}[n]|^2\Big{\}}={p_d}\mathbb{E}\bigg{\{}\Big{|}\sum\nolimits_{m=1}^{M}\sqrt{\eta_{mk'}}\textbf{g}_{mk}^{T}[n]\hat{\textbf{g}}_{mk'}^*[\lambda]\Big{|}^2\bigg{\}}\vspace{2 pt}\\&\displaystyle=p_d\mathbb{E}\Big{\{}\sum\nolimits_{m=1}^M\sum\nolimits_{m'=1}^M \sqrt{\eta_{mk'}\eta_{m'k'}}{\textbf{g}}_{mk}^T[n]\hat{\textbf{g}}_{mk'}^*[\lambda]\hat{\textbf{g}}_{m'k'}^T[\lambda]{\textbf{g}}_{m'k}^*[n]\Big{\}}.
\end{aligned}
\label{UI}
   \end{equation}
We decompose \eqref{UI} into two parts: $m=m'$ and $m\neq m'$. First, by utilizing \eqref{channel_estimate}, the procedure of $m=m'$ is shown as
\begin{equation}
\vspace{-3 pt}
\begin{array}{ll}
&\displaystyle\mathbb{E}\Big{\{}\sum\nolimits_{m=1}^M\eta_{mk'}{\textbf{g}}_{mk}^T[n]\hat{\textbf{g}}_{mk'}^*[\lambda]\hat{\textbf{g}}_{mk'}^T[\lambda]{\textbf{g}}_{mk}^*[n]\Big{\}}\vspace{3 pt }
\\&\displaystyle=\mathbb{E}\Big{\{}\sum\nolimits_{m=1}^M \eta_{mk}\rho_k^2[n-\lambda]\underbrace{{\textbf{g}}_{mk}^T[\lambda]\hat{\textbf{g}}_{mk'}^*[\lambda]\hat{\textbf{g}}_{mk'}^T[\lambda]{\textbf{g}}_{mk}^*[\lambda]}_{T_1}\Big{\}}
\\&\displaystyle+\mathbb{E}\Big{\{}\sum\nolimits_{m=1}^M \eta_{mk}\bar{\rho}_k^2[n-\lambda]\underbrace{{\textbf{e}}_{mk}^T[n]\hat{\textbf{g}}_{mk'}^*[\lambda]\hat{\textbf{g}}_{mk'}^T[\lambda]{\textbf{e}}_{mk}^*[n]}_{T_2}\Big{\}}.
\end{array}
\label{UI_part1}
   \end{equation}
        \begin{figure*}[t!]
           \vspace{-6 pt}
  \begin{equation}
\begin{array}{ll}
\displaystyle T_1 &\displaystyle={\textbf{g}}_{mk}^T[\lambda]\hat{\textbf{g}}_{mk'}^*[\lambda]\hat{\textbf{g}}_{mk'}^T[\lambda]{\textbf{g}}_{mk}^*[\lambda]={\textbf{g}}_{mk}^H[\lambda]\left(\hat{\textbf{g}}_{mk'}^\text{d}[\lambda]+\hat{\textbf{g}}_{mk'}^\text{c}[\lambda]\right)^H\left(\hat{\textbf{g}}_{mk'}^\text{d}[\lambda]+\hat{\textbf{g}}_{mk'}^\text{c}[\lambda]\right){\textbf{g}}_{mk}[\lambda]\\&\displaystyle=\text{tr}(\boldsymbol{\Delta}_{mk}\textbf{Q}_{mk'}^\text{c})+\text{tr}(\textbf{Q}_{mk'}^\text{d}\boldsymbol{\Delta}_{mk})+\underbrace{p_p\rho_k^2[\lambda-t_k^\text{d}]\text{tr}(\boldsymbol{\Delta}_{mk}^\text{d}\textbf{Z}_{mk',d})\text{tr}(\boldsymbol{\Delta}_{mk}^\text{d}\textbf{Z}^H_{mk',d})}_{k'\in\mathcal{P}_k^\text{d}}\\&\displaystyle+\underbrace{p_p\rho_k[\lambda-t_k^\text{d}]\rho_k[\lambda-t_k^\text{c}]\left(\text{tr}(\boldsymbol{\Delta}_{mk}^\text{d}\textbf{Z}_{mk',d})\text{tr}(\boldsymbol{\Delta}_{mk}^\text{c}\textbf{Z}^H_{mk',c})+\text{tr}(\boldsymbol{\Delta}_{mk}^\text{c}\textbf{Z}_{mk',c})\text{tr}(\boldsymbol{\Delta}_{mk}^\text{d}\textbf{Z}^H_{mk',d})\right)}_{k'\in\mathcal{P}_k^\text{d}\cap\mathcal{P}_k^\text{c}}\\&\displaystyle+\sum\nolimits_{k''\in\mathcal{P}_{k'}^c}p_p\sum\nolimits_{j=1}^J\beta_{mj}^2\beta_{kj}\beta_{k''j}\text{tr}(\textbf{T}_j^2)\text{tr}(\textbf{R}_{mj,r}\textbf{Z}_{mk',c})\text{tr}(\textbf{R}_{mj,r}\textbf{Z}^H_{mk',c})+\sum\nolimits_{j=1}^J\beta_{mj}^2\beta_{kj}\sigma^2_j\text{tr}(\textbf{T}_j^2)\text{tr}(\textbf{R}_{mj,r}\textbf{Z}_{mk',c})\text{tr}(\textbf{R}_{mj,r}\textbf{Z}^H_{mk',c})\\&\displaystyle+\underbrace{p_p\rho_k^2[\lambda-t_k^\text{c}]\left(\text{tr}(\boldsymbol{\Delta}_{mk}^\text{c}\textbf{Z}_{mk',c})\text{tr}(\boldsymbol{\Delta}_{mk}^\text{c}\textbf{Z}^H_{mk',c})+\sum\nolimits_{j=1}^J\beta_{mj}^2\beta_{kj}^2\text{tr}(\textbf{T}_j^2)\text{tr}(\textbf{R}_{mj,r}\textbf{Z}_{mk',c}\textbf{R}_{mj,r}\textbf{Z}^H_{mk',c})\right)}_{k'\in\mathcal{P}_k^c},
\end{array} \vspace{-3 pt}
\label{part_1}
   \end{equation}
   \hrulefill
   \end{figure*}
             \begin{figure*}[t!]
            % \vspace{-6 pt}
  \begin{equation}
\begin{array}{ll}
\displaystyle T_2 &\displaystyle={\textbf{e}}_{mk}^T[n]\hat{\textbf{g}}_{mk'}^*[\lambda]\hat{\textbf{g}}_{mk'}^T[\lambda]{\textbf{e}}_{mk}^*[n]={\textbf{e}}_{mk}^H[n]\left(\hat{\textbf{g}}_{mk'}^\text{d}[\lambda]+\hat{\textbf{g}}_{mk'}^\text{c}[\lambda]\right)^H\left(\hat{\textbf{g}}_{mk'}^\text{d}[\lambda]+\hat{\textbf{g}}_{mk'}^\text{c}[\lambda]\right){\textbf{e}}_{mk}[n]\\&\displaystyle=\text{tr}(\boldsymbol{\Delta}_{mk}\textbf{Q}_{mk'}^\text{c})+\text{tr}(\textbf{Q}_{mk'}^\text{d}\boldsymbol{\Delta}_{mk})\\&\displaystyle+\sum\nolimits_{k''\in\mathcal{P}_{k'}^c}p_p\sum\nolimits_{j=1}^J\beta_{mj}^2\beta_{kj}\beta_{k''j}\text{tr}(\textbf{T}_j^2)\text{tr}(\textbf{R}_{mj,r}\textbf{Z}_{mk',c})\text{tr}(\textbf{R}_{mj,r}\textbf{Z}^H_{mk',c})+\sum\nolimits_{j=1}^J\beta_{mj}^2\beta_{kj}\sigma^2_j\text{tr}(\textbf{T}_j^2)\text{tr}(\textbf{R}_{mj,r}\textbf{Z}_{mk',c})\text{tr}(\textbf{R}_{mj,r}\textbf{Z}^H_{mk',c}).
\end{array} \vspace{-3 pt}
\label{part_2}
   \end{equation}
   \hrulefill
   \end{figure*}      
        The relevant terms can be further described as
        \eqref{part_1} at the top of this page and \eqref{part_2} at the top of the next page.
Similarly, the procedure for $m\neq m'$ can be simplified as \eqref{UI_part2} at the top of the next page.
\begin{figure*}[t!]
\vspace{-8 pt}
\begin{equation}
\begin{array}{ll}
\displaystyle \mathbb{E}\Big{\{}\sum\nolimits_{m=1}^M\sum\nolimits_{m'=1\atop m'\neq m}^M \sqrt{\eta_{mk'}\eta_{m'k'}}{\textbf{g}}_{mk}^T[n]\hat{\textbf{g}}_{mk'}^*[\lambda]\hat{\textbf{g}}_{m'k'}^T[\lambda]{\textbf{g}}_{m'k}^*[n]\Big{\}}
&=\displaystyle\mathbb{E}\Big{\{}\sum\nolimits_{m=1}^M\sum\nolimits_{m'=1\atop m'\neq m}^M \sqrt{\eta_{mk'}\eta_{m'k'}}\rho_k^2[n-\lambda]\underbrace{{\textbf{g}}_{mk}^T[\lambda]\hat{\textbf{g}}_{mk'}^*[\lambda]\hat{\textbf{g}}_{m'k'}^T[\lambda]{\textbf{g}}_{m'k}^*[\lambda]}_{T_3}\Big{\}}\\&+\displaystyle\mathbb{E}\Big{\{}\sum\nolimits_{m=1}^M\sum\nolimits_{m'=1\atop m'\neq m}^M \sqrt{\eta_{mk'}\eta_{m'k'}}\bar{\rho}_k^2[n-\lambda]\underbrace{{\textbf{e}}_{mk}^T[n]\hat{\textbf{g}}_{mk'}^*[\lambda]\hat{\textbf{g}}_{m'k'}^T[\lambda]{\textbf{e}}_{m'k}^*[n]}_{T_4}\Big{\}},
\end{array} \vspace{-3 pt}
 \label{UI_part2}
   \end{equation}
   \hrulefill
   \end{figure*}
Then, we derive the relevant terms by \eqref{part_3} and \eqref{part_4} at the top of the next page.
           \begin{figure*}[t!]
           \vspace{-6 pt}
  \begin{equation}
\begin{array}{ll}
\displaystyle T_3 &\displaystyle={\textbf{g}}_{mk}^T[\lambda]\hat{\textbf{g}}_{mk'}^*[\lambda]\hat{\textbf{g}}_{m'k'}^T[\lambda]{\textbf{g}}_{m'k}^*[\lambda]={\textbf{g}}_{mk}^H[\lambda]\left(\hat{\textbf{g}}_{mk'}^\text{d}[\lambda]+\hat{\textbf{g}}_{mk'}^\text{c}[\lambda]\right)^H\left(\hat{\textbf{g}}_{m'k'}^\text{d}[\lambda]+\hat{\textbf{g}}_{m'k'}^\text{c}[\lambda]\right){\textbf{g}}_{m'k}[\lambda]\\&\displaystyle=\underbrace{p_p\rho_k^2[\lambda-t_k^\text{d}]\text{tr}(\boldsymbol{\Delta}_{mk}^\text{d}\textbf{Z}_{mk',d})\text{tr}(\boldsymbol{\Delta}_{m'k}^\text{d}\textbf{Z}^H_{m'k',d})}_{k'\in\mathcal{P}_k^\text{d}}+\underbrace{p_p\rho_k^2[\lambda-t_k^\text{c}]\text{tr}(\boldsymbol{\Delta}_{mk}^\text{c}\textbf{Z}_{mk',c})\text{tr}(\boldsymbol{\Delta}_{m'k}^\text{c}\textbf{Z}^H_{m'k',c})}_{k'\in\mathcal{P}_k^c}\\&\displaystyle+\underbrace{p_p\rho_k[\lambda-t_k^\text{d}]\rho_k[\lambda-t_k^\text{c}]\left(\text{tr}(\boldsymbol{\Delta}_{mk}^\text{d}\textbf{Z}_{mk',d})\text{tr}(\boldsymbol{\Delta}_{m'k}^\text{c}\textbf{Z}^H_{m'k',c})+\text{tr}(\boldsymbol{\Delta}_{mk}^\text{c}\textbf{Z}_{mk',c})\text{tr}(\boldsymbol{\Delta}_{m'k}^\text{d}\textbf{Z}^H_{m'k',d})\right)}_{k'\in\mathcal{P}_k^\text{d}\cap\mathcal{P}_k^\text{c}}\\&\displaystyle+\sum\limits_{k''\in\mathcal{P}_{k'}^c}p_p\sum\limits_{j=1}^J\beta_{mj}\beta_{m'j}\beta_{kj}\beta_{k''j}\text{tr}(\textbf{T}_j^2)\text{tr}(\textbf{R}_{mj,r}\textbf{Z}_{mk',c})\text{tr}(\textbf{R}_{m'j,r}\textbf{Z}^H_{m'k',c})+\sum\limits_{j=1}^J\beta_{mj}\beta_{m'j}\beta_{kj}\sigma^2_j\text{tr}(\textbf{T}_j^2)\text{tr}(\textbf{R}_{mj,r}\textbf{Z}_{mk',c})\text{tr}(\textbf{R}_{m'j,r}\textbf{Z}^H_{m'k',c}),
\end{array} \vspace{-3 pt}
\label{part_3}
   \end{equation}
   \hrulefill
   \end{figure*}
             \begin{figure*}[t!]
             \vspace{-8 pt}
  \begin{equation}
\begin{array}{ll}
\displaystyle T_4 &\displaystyle={\textbf{e}}_{mk}^T[n]\hat{\textbf{g}}_{mk'}^*[\lambda]\hat{\textbf{g}}_{m'k'}^T[\lambda]{\textbf{e}}_{m'k}^*[n]={\textbf{e}}_{mk}^H[n]\left(\hat{\textbf{g}}_{mk'}^\text{d}[\lambda]+\hat{\textbf{g}}_{mk'}^\text{c}[\lambda]\right)^H\left(\hat{\textbf{g}}_{m'k'}^\text{d}[\lambda]+\hat{\textbf{g}}_{m'k'}^\text{c}[\lambda]\right){\textbf{e}}_{m'k}[n]\\&\displaystyle=\sum\limits_{k''\in\mathcal{P}_{k'}^c}p_p\sum\limits_{j=1}^J\beta_{mj}\beta_{m'j}\beta_{kj}\beta_{k''j}\text{tr}(\textbf{T}_j^2)\text{tr}(\textbf{R}_{mj,r}\textbf{Z}_{mk',c})\text{tr}(\textbf{R}_{m'j,r}\textbf{Z}^H_{m'k',c})+\sum\limits_{j=1}^J\beta_{mj}\beta_{m'j}\beta_{kj}\sigma^2_j\text{tr}(\textbf{T}_j^2)\text{tr}(\textbf{R}_{mj,r}\textbf{Z}_{mk',c})\text{tr}(\textbf{R}_{m'j,r}\textbf{Z}^H_{m'k',c}).
\end{array} \vspace{-3 pt}
\label{part_4}
   \end{equation}
   \hrulefill
   \end{figure*}
According to the above procedure, the proof is completed.

\section
{Proof of Complex Gradients}
This appendix delivers the derivations of complex gradients $\nabla _{{\boldsymbol{\Phi}}_j}\text{SE}_{\text{sum}}({\boldsymbol{\Phi}})$ in \eqref{Optimization_4} regarding ${{\boldsymbol{\Phi}}}_j,~\forall j$. When considering $\text{tr}(\textbf{T}_{j})$, we can obtain 
\begin{equation}
    \begin{array}{ll}
\displaystyle
d\text{tr}(\textbf{T}_j)& \displaystyle=\text{tr}\Big{(}d\big{(}A_j^2{\textbf{R}_j}^{1/2}\boldsymbol{\Phi}_{j}\textbf{R}_j\boldsymbol{\Phi}_{j}^H{\textbf{R}_j}^{1/2}\big{)}\Big{)}\\ &\displaystyle=A_j^2\text{tr}\bigg{(}\textbf{R}_j{\boldsymbol{\Phi}}_{j}^H{\textbf{R}}_jd{\boldsymbol{\Phi}}_{j}+\left({\textbf{R}}_j{\boldsymbol{\Phi}}_{j}\textbf{R}_j\right)^Td{\boldsymbol{\Phi}}^*_{j}\bigg{)},
   \end{array}
   \label{derivative_pre}
\end{equation}
since ${{\boldsymbol{\Phi}}_j}$ is a diagonal matrix, then so must $d{{\boldsymbol{\Phi}}_j}$. With the help of \cite{Minka2000OldAN}, we can obtain
\begin{equation}
    \begin{array}{ll}
\displaystyle
\nabla _{{\boldsymbol{\Phi}}_{j}}\text{tr}(\textbf{T}_{j})\displaystyle=A_j^2\Big{(}\textbf{R}_j{\boldsymbol{\Phi}}_{j}^H{\textbf{R}}_j\Big{)}^T\circ\textbf{I}=A_j^2\text{diag}\Big{(}{\textbf{R}}_j^T{\boldsymbol{\Phi}}_{j}^*{\textbf{R}}_j^T\Big{)}.
   \end{array}
   \label{derivative_pre_1}
\end{equation}

Based on $\mathbf{\Delta}_{mk}^\text{c} $ in \eqref{Delta_c} and $\mathbf{\Psi}_{mk}^\text{c}$ in \eqref{Psi_c}, we can derive the differentials $d(\mathbf{\Delta}_{mk}^\text{c} )$ and $d((\mathbf{\Psi}_{mk}^\text{c} )^{-1})$ as follows. With respect to $d(\mathbf{\Delta}_{mk}^\text{c} )$, we can obtain \cite{Minka2000OldAN,4203075}
\begin{equation}
    \begin{array}{ll}
\displaystyle
d(\mathbf{\Delta}_{mk}^\text{c} )&\displaystyle=d\Big{(}\sum\nolimits_{j=1}^J\beta_{mj}\beta_{kj}\text{tr}(\textbf{T}_{j}){\textbf{R}}_{mj,r}\Big{)}
=\sum\nolimits_{j=1}^J\mathbf{\Pi}_{mkj}d\text{tr}(\textbf{T}_{j}) ,
   \end{array}
   \label{derivative}
\end{equation}
where $\mathbf{\Pi}_{mkj}=\beta_{mj}\beta_{kj}{\textbf{R}}_{mj,r}$.
By applying \eqref{derivative_pre}-\eqref{derivative_pre_1}, we can have
\begin{equation}
    \begin{array}{ll}
\displaystyle
\nabla _{{\boldsymbol{\Phi}}_{j}}\text{tr}(\mathbf{\Delta}_{mk}^\text{c} )
\displaystyle=A_j^2\beta_{mj}\beta_{kj}\text{tr}\Big{(}{\textbf{R}}_{mj,r}\Big{)}\text{diag}\Big{(}{\textbf{R}}_j^T{\boldsymbol{\Phi}}_{j}^*{\textbf{R}}_j^T\Big{)}.
   \end{array}
   \label{derivative_1}
\end{equation}
Referring to \cite{10297571}, we can derive $d\left((\mathbf{\Psi}_{mk}^\text{c})^{-1}\right)$ as
\begin{equation}
    \begin{array}{ll}
\displaystyle
d\left((\mathbf{\Psi}_{mk}^\text{c})^{-1}\right)=-(\mathbf{\Psi}_{mk}^\text{c})^{-1}d(\mathbf{\Psi}_{mk}^\text{c})(\mathbf{\Psi}_{mk}^\text{c})^{-1},
   \end{array}
   \label{derivative_4}
\end{equation}
where $d(\mathbf{\Psi}_{mk}^\text{c})$ can be obtained by %\eqref{derivative_5} at the top of the next page
%\begin{figure*}[t!]
\begin{equation}
    \begin{array}{ll}
\displaystyle
d(\mathbf{\Psi}_{mk}^\text{c})& \displaystyle=d\bigg{(}\sum\limits_{k'\in\mathcal{P}_k^\text{c}}p_p\mathbf{\Delta}_{mk'}^\text{c}+\displaystyle\sum\limits_{j=1}^J\beta_{mj}\sigma_j^2\text{tr}(\textbf{T}_j)
     \textbf{R}_{mj,r}+\sigma^2\textbf{I}_N\bigg{)}\\   & \displaystyle
     =\sum\limits_{k'\in\mathcal{P}_k^\text{c}}p_pd(\mathbf{\Delta}_{mk'}^\text{c})+\sum\limits_{j=1}^J\beta_{mj}\sigma_j^2d\left(\text{tr}(\textbf{T}_j)\right)
     \textbf{R}_{mj,r},
   \end{array}
   \label{derivative_5}
\end{equation}
   
   Then, we can obtain $d(\textbf{Q}_{mk}^\text{c})$ as \eqref{derivative_10} and the differential of $\text{tr}(\textbf{Q}_{mk}^\text{c})$ as \eqref{derivative_11} at the top of the next page \cite{10297571,4203075}. Next, we can obtain the derivative $\nabla _{{\boldsymbol{\Phi}}_{j}}\text{tr}(\textbf{Q}_{mk}^\text{c})$ as \eqref{derivative_12} at the top of this page.
   \begin{figure*}[t!]
   \vspace{-6 pt}
  \begin{equation}
    \begin{array}{ll}
\displaystyle d(\textbf{Q}_{mk}^\text{c})&  \displaystyle=d(\textbf{R}_{mk}^\text{c}(\mathbf{\Psi}_{mk}^\text{c})^{-1}\textbf{R}_{mk}^\text{c})=d(\textbf{R}_{mk}^\text{c})(\mathbf{\Psi}_{mk}^\text{c})^{-1}\textbf{R}_{mk}^\text{c}+\textbf{R}_{mk}^\text{c}d\left((\mathbf{\Psi}_{mk}^\text{c})^{-1}\right)\textbf{R}_{mk}^\text{c}+\textbf{R}_{mk}^\text{c}(\mathbf{\Psi}_{mk}^\text{c})^{-1}d(\textbf{R}_{mk}^\text{c})\\ &\displaystyle=d(\textbf{R}_{mk}^\text{c})(\mathbf{\Psi}_{mk}^\text{c})^{-1}\textbf{R}_{mk}^\text{c}-\textbf{R}_{mk}^\text{c}(\mathbf{\Psi}_{mk}^\text{c})^{-1}d(\mathbf{\Psi}_{mk}^\text{c})(\mathbf{\Psi}_{mk}^\text{c})^{-1}\textbf{R}_{mk}^\text{c}+\textbf{R}_{mk}^\text{c}(\mathbf{\Psi}_{mk}^\text{c})^{-1}d(\textbf{R}_{mk}^\text{c}),
  \end{array} \vspace{-3 pt}
   \label{derivative_10}
   \end{equation} 
   \hrulefill
      \end{figure*}
\begin{figure*}[t!]
\vspace{-6 pt}
  \begin{equation}
    \begin{array}{ll}
\displaystyle d\big{(}\text{tr}(\textbf{Q}_{mk}^\text{c})\big{)}&\displaystyle=\text{tr}\big{(}d(\textbf{Q}_{mk}^\text{c})\big{)}=\text{tr}\bigg{(}d(\textbf{R}_{mk}^\text{c})(\mathbf{\Psi}_{mk}^\text{c})^{-1}\textbf{R}_{mk}^\text{c}-\textbf{R}_{mk}^\text{c}(\mathbf{\Psi}_{mk}^\text{c})^{-1}d(\mathbf{\Psi}_{mk}^\text{c})(\mathbf{\Psi}_{mk}^\text{c})^{-1}\textbf{R}_{mk}^\text{c}+\textbf{R}_{mk}^\text{c}(\mathbf{\Psi}_{mk}^\text{c})^{-1}d(\textbf{R}_{mk}^\text{c})\bigg{)}\\&\displaystyle= p_p\rho_k^2[\lambda-t_k^\text{c}]\text{tr}\left(d(\mathbf{\Delta}_{mk}^\text{c})(\mathbf{\Psi}_{mk}^\text{c})^{-1}\mathbf{\Delta}_{mk}^\text{c}-\mathbf{\Delta}_{mk}^\text{c}(\mathbf{\Psi}_{mk}^\text{c})^{-1}d(\mathbf{\Psi}_{mk}^\text{c})(\mathbf{\Psi}_{mk}^\text{c})^{-1}\mathbf{\Delta}_{mk}^\text{c}+\mathbf{\Delta}_{mk}^\text{c}(\mathbf{\Psi}_{mk}^\text{c})^{-1}d(\mathbf{\Delta}_{mk}^\text{c})\right)\\&\displaystyle= p_p\rho_k^2[\lambda-t_k^\text{c}]\left[\begin{array}{ll}\displaystyle\sum\nolimits_{j=1}^J\text{tr}\left((\mathbf{\Psi}_{mk}^\text{c})^{-1}\mathbf{\Delta}_{mk}^\text{c}\mathbf{\Pi}_{mkj}\right)+\sum\nolimits_{j=1}^J\text{tr}\left(\mathbf{\Delta}_{mk}^\text{c}(\mathbf{\Psi}_{mk}^\text{c})^{-1}\mathbf{\Pi}_{mkj}\right)\\\displaystyle-\sum\nolimits_{j=1}^J\sum\nolimits_{k'\in\mathcal{P}_{k}^\text{c}}p_p\text{tr}\left((\mathbf{\Psi}_{mk}^\text{c})^{-1}\mathbf{\Delta}_{mk}^\text{c}\mathbf{\Delta}_{mk}^\text{c}(\mathbf{\Psi}_{mk}^\text{c})^{-1}\mathbf{\Pi}_{mk'j}\right)\\\displaystyle-\sum\nolimits_{j=1}^J\beta_{mj}\sigma_j^2\text{tr}\left((\mathbf{\Psi}_{mk}^\text{c})^{-1}\mathbf{\Delta}_{mk}^\text{c}\mathbf{\Delta}_{mk}^\text{c}(\mathbf{\Psi}_{mk}^\text{c})^{-1}\textbf{R}_{mj,r}\right)
\end{array}
\right]d\text{tr}(\textbf{T}_j)
  \end{array} \vspace{-3 pt}
   \label{derivative_11}
   \end{equation} 
   \hrulefill
\end{figure*}
\begin{figure*}[t!]
\vspace{-6 pt}
  \begin{equation}
    \begin{array}{ll}
\displaystyle\nabla _{{\boldsymbol{\Phi}}_{j}}\text{tr}(\textbf{Q}_{mk}^\text{c})&\displaystyle= A_j^2p_p\rho_k^2[\lambda-t_k^\text{c}]\left[\begin{array}{ll}\displaystyle\text{tr}\left((\mathbf{\Psi}_{mk}^\text{c})^{-1}\mathbf{\Delta}_{mk}^\text{c}\mathbf{\Pi}_{mkj}\right)+\text{tr}\left(\mathbf{\Delta}_{mk}^\text{c}(\mathbf{\Psi}_{mk}^\text{c})^{-1}\mathbf{\Pi}_{mkj}\right)\\\displaystyle-\sum\nolimits_{k'\in\mathcal{P}_{k}^\text{c}}p_p\text{tr}\left((\mathbf{\Psi}_{mk}^\text{c})^{-1}\mathbf{\Delta}_{mk}^\text{c}\mathbf{\Delta}_{mk}^\text{c}(\mathbf{\Psi}_{mk}^\text{c})^{-1}\mathbf{\Pi}_{mk'j}\right)\\\displaystyle-\beta_{mj}\sigma_j^2\text{tr}\left((\mathbf{\Psi}_{mk}^\text{c})^{-1}\mathbf{\Delta}_{mk}^\text{c}\mathbf{\Delta}_{mk}^\text{c}(\mathbf{\Psi}_{mk}^\text{c})^{-1}\textbf{R}_{mj,r}\right)
\end{array}
\right]\text{diag}\Big{(}{\textbf{R}}_j^T{\boldsymbol{\Phi}}_{j}^*{\textbf{R}}_j^T\Big{)},
  \end{array} \vspace{-3 pt}
   \label{derivative_12}
   \end{equation} 
   \hrulefill
\end{figure*}
Similarly,  we can obtain the derivative $\nabla _{{\boldsymbol{\Phi}}_{j}}\text{tr}(\textbf{Z}_{mk,c})$ as \eqref{derivative_13} at the top of the next page.
\begin{figure*}[t!]
\vspace{-6 pt}
  \begin{equation}
    \begin{array}{ll}
\displaystyle\nabla _{{\boldsymbol{\Phi}}_{j}}\text{tr}(\textbf{Z}_{mk,c})&\displaystyle= A_j^2\sqrt{p_p}\rho_k[\lambda-t_k^\text{c}]\left[\begin{array}{ll}\text{tr}\left((\mathbf{\Psi}_{mk}^\text{c})^{-1}\mathbf{\Pi}_{mkj}\right)-\displaystyle\sum\limits_{k'\in\mathcal{P}_{k}^\text{c}}p_p\text{tr}\left((\mathbf{\Psi}_{mk}^\text{c})^{-1}\mathbf{\Delta}_{mk}^\text{c}\mathbf{\Psi}_{mk}^\text{c})^{-1}\mathbf{\Pi}_{mk'j}\right)\\-\beta_{mj}\sigma_j^2\text{tr}\left((\mathbf{\Psi}_{mk}^\text{c})^{-1}\mathbf{\Delta}_{mk}^\text{c}(\mathbf{\Psi}_{mk}^\text{c})^{-1}\textbf{R}_{mj,r}\right)
\end{array}
\right]\text{diag}\Big{(}{\textbf{R}}_j^T{\boldsymbol{\Phi}}_{j}^*{\textbf{R}}_j^T\Big{)}.
  \end{array}
   \label{derivative_13}
   \end{equation} 
   \hrulefill
\end{figure*}

\end{appendices}

% or


%\appendices

%\section{}

%\section*{Acknowledgment}

\ifCLASSOPTIONcaptionsoff
  \newpage
\fi



\bibliographystyle{IEEEtran}
\bibliography{IEEEabrv,ref.bib}
\end{document}